\documentclass[12pt]{article}

\usepackage[margin=1in]{geometry}  
\usepackage{epsfig}
\usepackage{graphics,graphicx,color}
\usepackage{amssymb,amsfonts,amsthm,amsmath}
\usepackage{multirow}
\usepackage{mathrsfs}
\usepackage{natbib}
\usepackage{enumerate}

\newtheorem{theorem}{Theorem}

\newcommand{\bm}[1]{\mbox{\boldmath$ #1 $\unboldmath}}  

\begin{document}

\begin{center}
{\Large\bf Space-Filling Designs for Robustness Experiments}\\
\hfill \\
V. Roshan Joseph\\
{\small H. Milton Stewart School of Industrial and Systems
Engineering,} \\ {\small Georgia Institute of
Technology,}
{\small Atlanta, GA 30332-0205}\\
\hfill \\
Li Gu\\
    {\small Wells Fargo \& Company,} \\
    {\small Charlotte, NC 28202} \\
    \hfill \\

Shan Ba and William R. Myers\\
{\small Quantitative Sciences}\\ {\small The Procter \& Gamble Company,}
{\small Mason, OH 45040}\\

\end{center}

\begin{abstract}
To identify the robust settings of the control factors, it is very important to understand how they interact with the noise factors. In this article, we propose space-filling designs for computer experiments that are more capable of accurately estimating the control-by-noise interactions. Moreover, the existing space-filling designs focus on uniformly distributing the points in the design space, which are not suitable for noise factors because they usually follow non-uniform distributions such as normal distribution. This would suggest placing more points in the regions with high probability mass. However, noise factors also tend to have a smooth relationship with the response and therefore, placing more points towards the tails of the distribution is also useful for accurately estimating the relationship. These two opposing effects make the experimental design methodology a challenging problem. We propose optimal and computationally efficient solutions to this problem and demonstrate their advantages using simulated examples and a real industry example involving a manufacturing packing line.
\end{abstract}

\begin{quote}
\end{quote}

\section{INTRODUCTION}
\label{3sec:intro}

Robust parameter design is a cost-efficient technique for quality improvement. Originally proposed by \citet{Taguchi1987}, the technique has been widely adopted in industries for system (product or process) optimization. The core idea is to first divide the factors in the system into two groups: control factors and noise factors. Control factors are those factors in the system than can be cost-effectively controlled. On the other hand, noise factors are those factors which are either impossible or too expensive to control. For example, in product design of a razor for shaving, blade thickness, gap between the blades, angle of the blades, etc. are control factors, whereas consumer attributes like the skin type, hair length and density, and product usage attributes like the pressure applied on the skin, handle angle, etc. are noise factors. Since the noise factors are uncontrollable, they introduce variability in the performance of the product. Robust parameter design is a technique to find a setting of the control factors (also known as parameter design) that will make the system robust or insensitive to the noise factors. Thus, under a robust parameter design, the output becomes less affected by the noise variability even when the noise factors are left uncontrolled. This is why the approach using robust parameter design is less costly than the other quality improvement techniques, which try to directly control the noise factors in the system.

The key to a successful robust parameter design is in identifying important control-by-noise interactions of the system. Only when such interactions exist we can use the control factors to reduce the sensitivity of the noise factors. These interactions are usually unknown in practice and their existence need to be investigated through experimentation. Thus designing good experiments is a crucial step in robustness studies. Many efficient experimental design techniques are proposed in the literature such as cross arrays \citep{Taguchi1987} and single arrays \citep{Welch1990, Shoemaker1991, Wu2003, Kang2009}. A thorough discussion of these techniques can be found in the books by \citet{Wu2009} or \citet{Myers2016}.

The aforementioned experimental design techniques are mainly proposed for physical experimentation except for the work of \cite{Welch1990}. Recently computer experiments have become very common in industry. That is, if a computer model is available that can simulate the physical system, then the experiments can be performed in computers instead of the physical system. It is becoming very common for industry to develop a computer model for product design like simulating the performance of a razor. One particular example from Procter \& Gamble involving the development of a computer model that simulates a critical transformation of a packing line, that involves both control and noise factors, will be discussed in more detail later in the paper. Computer experiments can bring in tremendous cost savings because direct experimentation with the real physical system is always more expensive than investing on some computer time. However, there are several aspects of computer experiments that necessitate the use of a different experimental design technique or philosophy compared to those of physical experiments \citep{Sacks1989a}. Since most computer models are deterministic in nature, randomization and replications are not needed. Fractional factorial and orthogonal array-based design techniques that are prevalent in physical experiments lead to replications when projected onto subspace of factors and thus are unsuitable for computer experiments. Split-plot designs that are considered to be useful in robustness studies \citep{Bingham2003} become unnecessary as run orders and restrictions on randomization will not affect the computer model outputs. This led to the development of space-filling designs in computer experiments.

The existing work on robust parameter design using space-filling designs do not make any distinction between control and noise factors. A distinction is made only at the analysis stage \citep{Welch1990, Chen2006, Apley2006, Bates2006, Tan2015}. Sequential designs that directly attempt to find robust settings of control factors using expected improvement-type algorithms are proposed in the literature \citep{Williams2000, Lehman2004}, but we are not aware of any work on space-filling designs. It is important to develop space-filling designs that distinguish control and noise factors because their distributional properties are entirely different. Control factors are assumed to follow a uniform distribution, whereas noise factors typically follow non-uniform distributions such as normal distribution. Their nature of randomness is also different. Noise factors are intrinsically random and can vary over time and space. On the other hand, control factors remain fixed once their levels are chosen. A uniform distribution is imposed on the control factors only to represent our indifference on the choice of level given the range of possible values for each control factor.  Thus, unlike the control factors, most of the ``action'' in the noise factor space takes place in the regions of high probability mass. Therefore, space-filling designs that uniformly spread out points in the experimental region are not adequate for robust parameter design experiments. Moreover, the existing space-filling designs are not designed for precise estimation of control-by-noise interactions. In this article we propose a new version of space-filling design that is capable of estimating the control-by-noise interactions more precisely and puts more points in regions that matters the most.

The article is organized as follows. In Section 2, we propose a model-based optimal experimental design for robustness studies. Because of certain computational and practical difficulties associated with this approach, in Section 3, we propose a modified space-filling design as an alternative. In Section 4, we investigate the optimal choice of noise levels for the experiment. Extension of designs to deal with internal noise factors is proposed in Section 5. The proposed methodology is applied in a simulated example and the packing line computer experiment from Procter \& Gamble in Section 6. We conclude with some remarks in Section 7.

\section{MODEL-BASED OPTIMAL DESIGNS}
\label{3sec:design}

Let $\bm x=(x_1,\ldots,x_p)'$ be the set of control factors and $\bm z=(z_1,\ldots,z_q)'$ the set of (external) noise factors.  The case of internal noise factors and the differences between the two types will be discussed later. We assume that $\bm x \in \mathcal{X}=[0,1]^p$ and $\bm z \in \mathcal{Z}$, the support of the distribution of $\bm z$ which could be $\mathbb{R}^q$. The response $y$ is a deterministic function of both control and noise factors given by $y=g(\bm x,\bm z)$. Depending on the type of characteristic such as smaller-the-better, larger-the-better, or nominal-the-best, we can impose a quality loss function on $y$. Let $L(y)$ be such a loss function. Then, the objective of robust parameter design is to find the setting of control factors that minimizes the expected loss, where the expectation is taken with respect to the distribution of noise factors. Let $f(\bm z)$ denote the probability density function of $\bm z$. Then, the robust parameter design can be obtained by
\begin{equation}\label{eq:rpd0}
\min_{\bm x \in \mathcal{X}} \int_{\mathcal{Z}} L\{g(\bm x,\bm z)\}f(\bm z)d\bm z.
\end{equation}
Since the function $g(\cdot,\cdot)$ is available only as a computer code, an experiment will be conducted to estimate it. Let $\bm D=\{\bm x_1,\ldots,\bm x_p, \bm z_1,\ldots,\bm z_q\}$ be the experimental design with $n$ runs, where $\bm x_j=(x_{1j},\ldots,x_{nj})'$ and $\bm z_k=(z_{1k},\ldots,z_{nk})'$ denote the settings of the $j$th and $k$th control and noise factors, respectively. Let $y_i$  be the $i$th output from the computer model, $i=1,\ldots,n$.

A Gaussian process model or kriging \citep{Santner2003} is commonly used for estimating $g(\cdot,\cdot)$. So assume
\begin{equation}\label{eq:GP}
g(\bm x,\bm z)\sim GP(\mu,\tau^2 R(\cdot,\cdot)),
\end{equation}
where $\mu$ and $\tau^2$ are the unknown mean and variance parameters, and $R(\cdot,\cdot)$ is the correlation function. A commonly used correlation function is the Gaussian correlation function given by
\[R(\bm x_i-\bm x_j, \bm z_i-\bm z_j)=\exp\{-\sum_{l=1}^p\theta^x_l(x_{il}-x_{jl})^2-\sum_{l=1}^q\theta^z_l(z_{il}-z_{jl})^2\},\]
where $\bm \theta^x=(\theta^x_1,\ldots,\theta^x_p)'$ and $\bm \theta^z=(\theta^z_1,\ldots,\theta^z_q)'$ are the unknown correlation parameters of the control and noise factors, respectively. Let $\bm \theta$ be a $p+q$ column vector containing $\bm \theta^x$ and $\bm \theta^z$. We will use this correlation function throughout this article, but other correlation functions are also allowed as long as they can produce smooth realizations of the response in the noise factor space. The smoothness assumption with respect to noise factors is critical for our methodology and we will exploit it for developing the experimental designs. The Gaussian process can be viewed as a prior on the unknown function and therefore, we can obtain its posterior distribution using Bayes theorem \citep{Santner2003}:
\begin{equation}
g(\bm x,\bm z)|\bm y\sim N\left(\hat{g}(\bm x,\bm z), \tau^2MSE(\bm x,\bm z; \bm D, \bm \theta)\right),
\end{equation}
where $\hat{g}(\bm x,\bm z)=\mu+r(\bm x, \bm z)'\bm R^{-1}(\bm y-\mu \bm 1)$ and
\[MSE(\bm x,\bm z; \bm D, \bm \theta)=1-\bm r(\bm x, \bm z)'\bm R^{-1}\bm r(\bm x, \bm z),\]
where $\bm y=(y_1,\ldots,y_n)'$, $\bm r(\bm x, \bm z)$ is an $n\times 1$ vector with $i$th element $R(\bm x-\bm x_i, \bm z-\bm z_i)$, $\bm R$ is an $n\times n$ matrix with $ij$th element $R(\bm x_i-\bm x_j, \bm z_i-\bm z_j)$, and $\bm 1$ is a vector of 1's having length $n$. For simplicity, we chose to ignore the extra variability due to the estimation of $\mu$ and $\bm \theta$. See \cite{Muller2012} and \cite{Muller2014} for a discussion on the effect of this extra variability on experimental design.

The posterior mean $\hat{g}(\bm x,\bm z)$ can be used as an estimate of the response function from the experiment (also known as metamodel, surrogate model, or emulator). Then, the optimization in (\ref{eq:rpd0}) can be simplified as
\begin{equation}\label{eq:rpd}
\min_{\bm x \in \mathcal{X}} \int_\mathcal{Z} L\{\hat{g}(\bm x,\bm z)\}f(\bm z)d\bm z.
\end{equation}
It is also possible to incorporate the uncertainties in the estimation of $g(\cdot,\cdot)$ in the optimization as in \citet{Apley2011} and \citet{Tan2012}, but it will not be considered here for the sake of simplicity. The problem we are trying to solve is how to design the experiment $\bm D$ so that we can accurately estimate the solution to the optimization problem in (\ref{eq:rpd}).

Clearly, the optimization in (\ref{eq:rpd}) will give the true robust setting if $\hat{g}(\bm x,\bm z)$ is the true response surface, that is, if $MSE(\bm x,\bm z; \bm D, \bm \theta)=0$ for all $\bm x\in \mathcal{X}$ and $\bm z\in \mathcal{Z}$. Thus we should design the experiment so that $MSE(\bm x,\bm z; \bm D, \bm \theta)$ is as small as possible. Furthermore, a careful examination of (\ref{eq:rpd}) reveals an important insight on the experimental design problem. We need an accurate $g(\cdot,\cdot)$ only in the regions of $\bm z$ where $f(\bm z)$ is large. In other words, if $f(\bm z)$ is small in some regions, then the inaccuracies in the estimation of $g(\cdot,\cdot)$ in those regions will not affect the robust settings. This makes the experimental design problem for robustness different from that of a usual computer experiment. In fact, it makes sense to focus on the estimated solution to the optimization problem in (\ref{eq:rpd}) as proposed in \cite{Gins2006} rather than the estimation of $g(\cdot,\cdot)$. However, their approach works only for linear models fitted to physical experimental data. In contrast, the models considered in computer experiments are highly nonlinear and thus, finding an explicit solution to (\ref{eq:rpd}) is not feasible. So in this work we will focus on the estimation of $g(\cdot,\cdot)$.  Although this approach may not look ideal for the robustness objective, it does have certain advantages. The loss functions are many times loosely defined and one may want to investigate solutions to different possible choices of loss function \citep{Joseph2004}. Moreover, in real problems, there can be multiple quality characteristics and thus one may need to be satisfied with a compromise solution, which can be different from the optimal solution in (\ref{eq:rpd}). Thus an accurate $g(\cdot,\cdot)$ in the region of interest can be more beneficial than an accurate solution to (\ref{eq:rpd}) obtained for a specific choice of loss function and quality characteristic.

Thus, our aim is to find $\bm D$ such that $MSE(\bm x,\bm z; \bm D, \bm \theta)$ is small. However, since $MSE(\bm x,\bm z; \bm D, \bm \theta)$ is a function of $\bm x$ and $\bm z$, it is not possible to find such a design over the entire experimental region. Instead, a feasible approach is to minimize the average of $MSE(\bm x,\bm z; \bm D, \bm \theta)$, that is
\begin{equation}\label{eq:IMSE0}
\min_{\bm D}\int_\mathcal{X} \int_\mathcal{Z} MSE(\bm x,\bm z; \bm D, \bm \theta) f(\bm z) d\bm z d\bm x.
\end{equation}
This design criterion is the same as the  integrated mean squared error criterion in the literature \citep{Sacks1989a, Santner2003} except that we use the density of $\bm z$ as a weight function. This is quite a natural modification of the existing criterion and agrees with our intuition that we should give more weights for regions where $f(\bm z)$ is large. Surprisingly, we found that the solution to (\ref{eq:IMSE0}) places points in extremely low probability regions which are not very useful for finding the robust setting.  This problem can be alleviated if we use  the root-mean squared prediction error, which directly corresponds to confidence intervals of the prediction. Thus, consider a modified criterion
\begin{equation}\label{eq:IRMSE}
\min_{\bm D} IRMSE (\bm D, \bm \theta)=\min_{\bm D}\int_\mathcal{X} \int_\mathcal{Z} \sqrt{MSE(\bm x,\bm z; \bm D, \bm \theta)} f(\bm z) d\bm z d\bm x.
\end{equation}
This is a more meaningful criterion as it tries to minimize the expected volume of the confidence region of the predictions. We may generalize this criterion to
\begin{equation}\label{eq:IRMSEk}
\min_{\bm D} IRMSE_k (\bm D, \bm \theta)=\min_{\bm D}\left[\int_\mathcal{X} \int_\mathcal{Z} \left\{\sqrt{MSE(\bm x,\bm z; \bm D, \bm \theta)} f(\bm z)\right\}^k/C_k d\bm z d\bm x\right]^{1/k}
\end{equation}
for $k>0$ and $C_k=\int_{\mathcal{Z}}f^k(\bm z) d\bm z$. The special case of $k=2$ is of great interest as it is analytically tractable in some situations. Let $IMSE=IRMSE_2^2$. Thus,
\begin{equation}\label{eq:IMSE}
\min_{\bm D} IMSE (\bm D, \bm \theta)=\min_{\bm D}\int_\mathcal{X} \int_\mathcal{Z} MSE(\bm x,\bm z; \bm D, \bm \theta) f^2(\bm z)/C_2 d\bm z d\bm x.
\end{equation}
Interestingly, this is the same as the integrated mean squared error criterion in the literature but with a weight function $f^2(\bm z)$. In the case of uniform distributions,
$f(\cdot)$ or $f^2(\cdot)$ doesn't make any difference, but for non-uniform distribution this does
 make a big difference. We will see later that $f^2(\cdot)$ gives the right scaling and provides meaningful solutions to the robust parameter design problem.

A major challenge of using the foregoing criteria is that they are functions of the \emph{unknown} correlation parameters $\bm \theta$. One can minimize IRMSE (or IMSE) for a guessed value of $\bm \theta$, but the optimal design may not work well for another value of $\bm \theta$. A potential fix to overcome this problem is to first average the IRMSE over a prior distribution of $\bm \theta$ and find the design using
\begin{equation*}
\min_{\bm D}\int IRMSE(\bm D,\bm \theta)  p(\bm \theta)d\bm \theta.
\end{equation*}
This is a computationally intensive problem because $\bm R$ is a function of $\bm \theta$ and thus, inverting $\bm R$ and then integrating the IRMSE is time consuming. Moreover, this criterion doesn't work well in practice because $MSE(\bm x,\bm z;\bm D,\bm \theta)$ increases with $\bm \theta$ and therefore, the criterion is dominated by the large values of $\bm \theta$. \citet{Sacks1989b} proposed to overcome this problem by standardizing the criteria with respect to the optimal design obtained for a given value of $\bm \theta$. See \citet{Pratola2016} for a Bayesian version of this approach. Let $\Theta$ be a compact set containing the possible values of $\bm \theta$. Then, Sacks et al.'s approach is to find the design to maximize the minimum efficiency:
\begin{equation}\label{eq:RIRMSE}
\max_{\bm D} \min_{\bm \theta \in \Theta}\frac{IRMSE(\bm D^*(\bm\theta),\bm\theta) }{IRMSE(\bm D, \bm \theta)},
\end{equation}
where $\bm D^*(\bm \theta)=arg\min_{\bm D}IRMSE(\bm D, \bm \theta)$. This criterion is extremely computationally intensive because one needs to find the optimal design for every possible value of $\bm \theta \in \Theta$, which is difficult in high dimensions. \citet{Sacks1989b} tried to circumvent this problem by letting $\theta_i^x=\theta_j^z=\theta_0$ for all $i=1,\ldots,p$ and $j=1,\ldots,q$ and then taking a few discrete values of $\theta_0$. In our experience, this simplification results in designs having poor projections in subspaces of the factors, which is undesirable. One possible approach to improve the projections is to assign independent prior distributions for each of the unknown correlation parameters as in \cite{Joseph2015b}. However, this would make the computation of (\ref{eq:RIRMSE}) very expensive. For example, even if we choose only three discrete values for each correlation parameter, the total number of possible values in $\Theta$ would become $3^{p+q}$ which can be quite high when $p$ and/or $q$ is large.

Because of the foregoing difficulties, in the next section we will develop space-filling designs which are easy to compute and are model-robust. We will modify them so that they will perform well according to the criterion in (\ref{eq:RIRMSE}).

\section{SPACE-FILLING DESIGNS}
Space-filling designs aim at filling the experimental region evenly with as few gaps as possible. These designs are robust to modeling choices and thus, are widely used as designs for computer experiments.  Popular space-filling designs include Latin hypercube designs \citep{McKay1979}, distance-based designs such as maximin and minimax \citep{Johnson1990}, uniform designs \citep{Fang1994}, and several useful variants of them such as maximin Latin hypercube designs \citep{Morris1995} and maximum projection designs \citep{Joseph2015b}.  See \citet{Joseph2016} for a recent review of space-filling designs. However, these designs are developed for general purpose applications such as function approximation and not specifically for robustness experiments. As mentioned in the introduction, control-by-noise interactions are especially important for identifying robust settings. Therefore, we may hope to improve the performance of space-filling designs by improving their ability to estimate the control-by-noise interactions, possibly by sacrificing other not so important effects.

In the physical experiments' literature, there are mainly two classes of designs suitable for robustness experiments: cross arrays \citep{Taguchi1987} and single arrays \citep{Welch1990, Shoemaker1991}. To develop single arrays, one needs to first quantify the importance of each effect \citep{Bingham2003, Wu2003}. This was not too difficult with the fractional factorial experiments because such designs usually have only two or three levels for each factor. In contrast, computer experiments have large number of levels for each factor and therefore, numerous effects are involved in the modeling. This makes the effect ordering a difficult task. Thus, cross arrays seems to be an easier and straightforward approach for computer experiments.

To develop cross arrays, we first need to choose an $n_1$-run design $\bm D_x$ for the control factors (known as control array) and another $n_2$-run design $\bm D_z$ for the noise factors (known as noise array). Cross array can then be obtained by repeating the noise array for each run of the control array, which will have a total of $n_1n_2$ number of runs. We will denote the cross array by $\bm D=\bm D_x \times\bm D_z$. The suitability of a cross array for robustness experiments should be evident from its construction. We can estimate the effect of noise factors under each settings of the control factors in $\bm D_x$, which enables one to choose the control factor setting that makes the noise factors' effect on the response as small as possible. As shown in \cite{Wu2009} (see Theorem 11.1), if control factor effects are estimable from $\bm D_x$ and noise factor effects from $\bm D_z$, then the two-factor interactions between control and noise factors are estimable and clear in $\bm D$.

The following result shows how to construct an optimal cross array that minimizes the IMSE criterion in (\ref{eq:IMSE}). The proof is given in the Appendix.
\begin{theorem}
If a product correlation is used between control and noise factors, then an IMSE-optimal cross array can be obtained by crossing an IMSE-optimal control array and an IMSE-optimal noise array.
\end{theorem}
The IMSE optimal designs are closely related to the space-filling designs. For example, when $f(\bm z)=1$, as $k\rightarrow \infty$ in (\ref{eq:IRMSEk}),
\[IRMSE_{\infty}(\bm D)=\max_{\mathcal{X}}\max_{\mathcal{Z}}MSE(\bm x,\bm z;\bm D),\]
which is minimized by a minimax distance design when the correlations are small (Johnson et al. 1990). The assumption of small correlations is justifiable in the control factor space, but not in the noise factor space. Moreover, Noise factors have nonuniform distributions. We will discuss on how to modify the space-filling designs for the noise array in the next section.

\begin{figure}[h]
\begin{center}
\includegraphics[width = 0.75\textwidth]{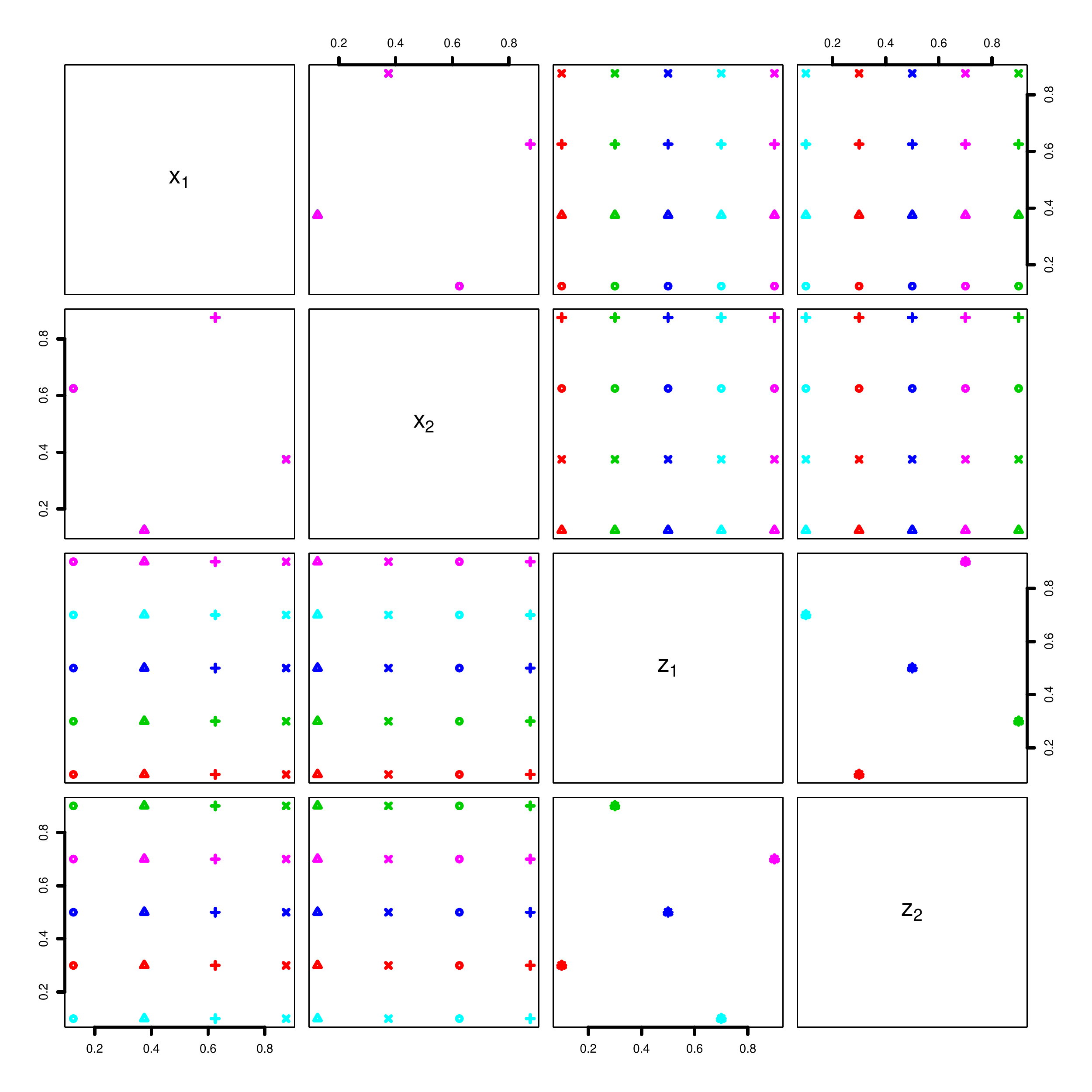}
\caption{Two-dimensional projections of a cross array obtained by crossing a 4-run MmLHD for control factors and 5-run MmLHD for noise factors.}
\label{fig:CA}
\end{center}
\end{figure}

A simple example can be used to illustrate why the cross arrays are useful in estimating control-by-noise interactions. Consider two control factors ($x_1$ and $x_2$) and two noise factors ($z_1$ and $z_2$). Suppose we choose a Maximin Latin hypercube design (MmLHD) with four runs for the control array and another MmLHD with five runs for the noise array. Then the cross array will have 20 runs. Their two-factor projections are shown in Figure \ref{fig:CA}. We can see that the projections are very poor in the control factor space and the noise factor space, but the projections are excellent on the control-by-noise factor space. Thus, we can obtain a good estimation of the control-by-noise interactions using this design. However, the estimation of the pure control or noise factor effects can be poor compared to using a 20-run MmLHD for the four factors.

\graphicspath{{./Figures/}}

Although the cross array looks promising for robustness experiments, it has certain disadvantages for using in computer experiments. Each control factor level is replicated $n_2$ number of times and noise factor level $n_1$ number of times. Thus, the number of levels is much smaller than that of a comparable single array, which has $n_1n_2$ number of levels for each factor. This can lead to poor estimation of nonlinear effects and higher order interactions in the control and noise factor spaces. This disadvantage of the cross array is amplified if there are only a few factors that are active. We propose an idea to overcome this disadvantage.

We can jitter each point in the cross array to increase the number of levels for each factor. The resulting design will still posses the good estimation ability of control-by-noise interactions because the response values observed over adjacent points are expected to be highly correlated. However, if the jittering radius is small, then there will be no improvement in the estimation of the pure control and noise factor effects. On the other hand, if the jittering radius is large, then the resulting design can lose its ability to efficiently estimate the control-by-noise interactions. So how much to jitter is a critical question and we have an intuitive solution for this. Let $r$ be the fill distance or covering radius of the cross array, which is defined as
\[r=\max_{\bm u \in [0,1]^{p+q}} ||\bm u-Q(\bm u,\bm D)||,\]
where $Q(\bm u,\bm D)$ is the closest point in $\bm D$ to $\bm u$ and $||\cdot||$ denotes the Euclidean distance. Let $B_i$ denote the ball with center at the $i$th design point $\bm D_i$ and radius $r$ and let $C_i$ be the hypercube inscribed in the ball. Now we can jitter the $i$th point within $C_i$. We chose $C_i$ instead of $B_i$ because in high dimensions the projected points inside the ball can be far away from the center. We will call the resulting design a \emph{Jittered Cross Array} (JCA).

We can do better than a random jittering. Since our aim is to overcome the issue with projections, we can choose the points in $C_i$ that will ensure good projections. We propose a sequential algorithm for doing this. Start from the center point $(.5,\ldots,.5)$ after rearranging the rows in the cross array so that the first run is closest to the center point. Now add one point from $C_i$ sequentially using the maximum projection (MaxPro) criterion \citep{Joseph2015b}:
\begin{equation}\label{eq:maxpro}
\bm D_{i}=\min_{\bm u \in \mathcal{C}_i}\sum_{j=1}^{i-1}\frac{1}{\prod_{l=1}^{p}{|u_{l}-\bm D_{jl}|^s}},
\end{equation}
for $i=2,\ldots,n$ and $s=2$. This algorithm adds points sequentially in a greedy manner such that the $i$th point is as far as possible from the previously chosen points under the MaxPro criterion. The algorithm is very easy to implement, but can converge to a local optimum. To improve the performance, we repeat this procedure many times by randomizing the order in which the $C_i$'s are chosen. Although the MaxPro criterion ensures that no two levels can be the same, the levels may not be equally spaced. So at the end of each iteration, we force them to be equally spaced, which can be easily done by ordering the levels. So the final design is like a Latin hypercube design (LHD), but with some clustering in the control and noise factor spaces. JCA shouldn't be confused with a cascading LHD \citep{Handcock2007}, which has clusters in the full-dimensional space and doesn't have a crossed array structure.

\begin{figure}[h]
\begin{center}
\includegraphics[width = 0.75\textwidth]{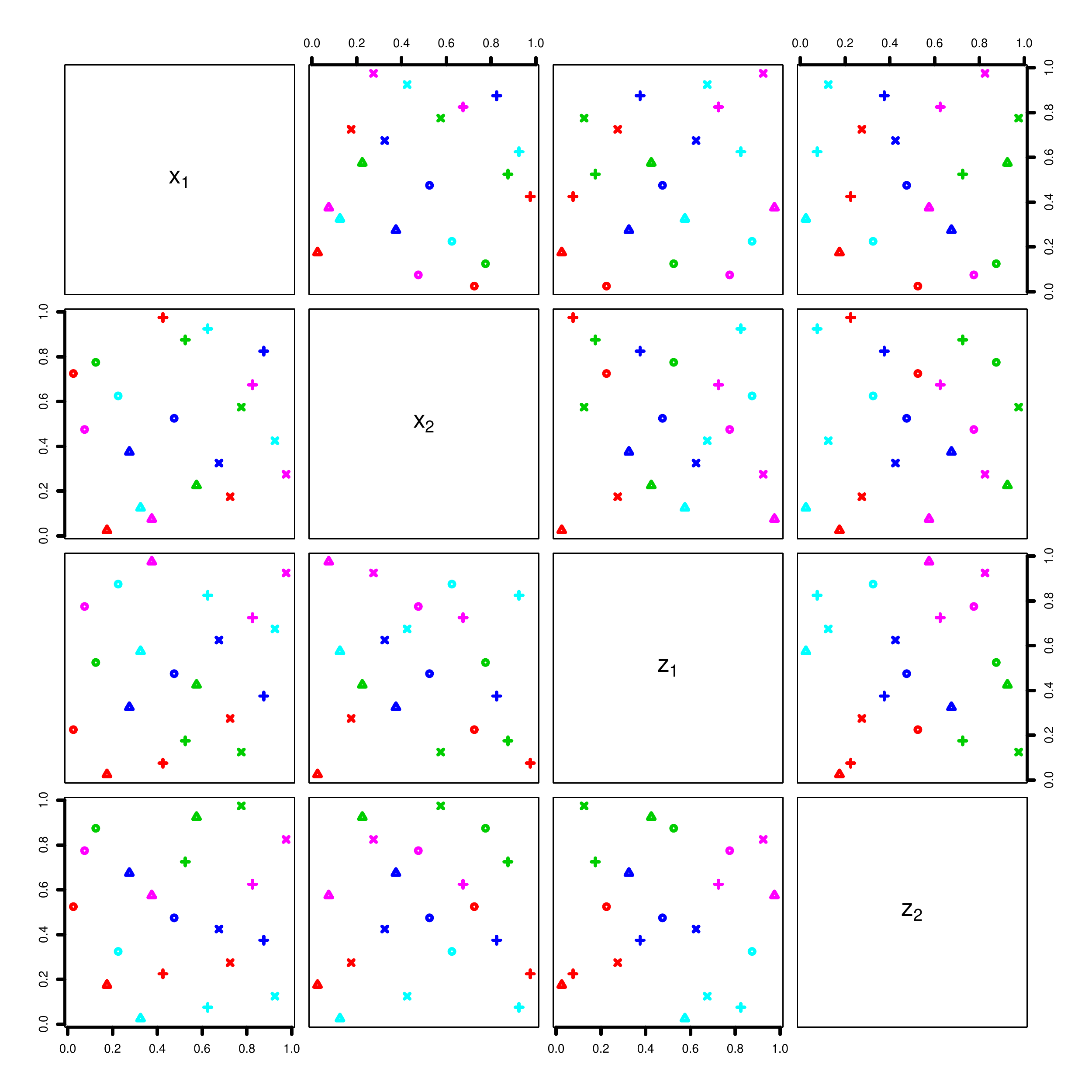}
\caption{Two-dimensional projections of the jittered cross array in 20 runs with two control and two noise factors. The control factor levels in the cross array are coded by the plotting symbols and noise factor levels by color.}
\label{fig:JCA}
\end{center}
\end{figure}

The two-dimensional projections of the JCA for the previous example is shown in Figure \ref{fig:JCA}. We can now see the 20 points in the $x_1\times x_2$ and $z_1\times z_2$ projections as opposed to only four and five points in the cross array. The projections in the control-by-noise spaces are still very good. We can also observe the clusters in the control (symbols) and noise (color) spaces, which shows that the cross array structure is approximately maintained. We will study the performance of these designs using simulated and real examples in a later section after deciding the optimal choice of noise array.
\graphicspath{{./Figures/}}

\section{NOISE ARRAY}

In this section we will discuss three possibilities for the choice of noise array. The first one is the most intuitive choice, but the latter two are better for robustness experiments.

\subsection{Transformed Design}
As mentioned earlier, the distribution of the noise factors are usually nonuniform. On the other hand, most space-filling designs are closely related to a uniform distribution as they try to spread out the points evenly in the experimental region. This suggests that we can possibly use the inverse probability transform method to transform the noise factor columns in a space-filling design to have the right distribution. More specifically, assume that the noise factors are independent. Let $F_l(z_l)$ be the distribution function of $z_l$, $l=1,\ldots,q$. Then it can be shown that if the design $\{\bm z_1,\ldots,\bm z_q\}$ minimizes the discrepancy from a uniform distribution, then $\{F_1^{-1}(\bm z_1),\ldots,F_q^{-1}(\bm z_q)\}$ minimizes the $F$-discrepancy \citep[p. 21]{Fang1994}, where the transformation is applied element-wise. We will call this design a \emph{transformed design}.

The independence assumption is crucial for the above simplification. This assumption can be easily relaxed when the noise factors follows a multivariate normal distribution: $N(\bm 0,\bm \Sigma_z)$. We can first find the space-filling design assuming independence and then transform using $\bm \Sigma^{1/2}\Phi^{-1}(\bm z^i)$, where $\Phi(\cdot)$ is the distribution function of a standard normal variable and $\bm z^i$ is the $i$th row of the space-filling design. If the distribution is not normal, then the design can be found using the idea of support points \citep{Mak2017}.

\begin{figure}[h]
\begin{center}
\begin{tabular}{cc}
\includegraphics[width = 0.45\textwidth]{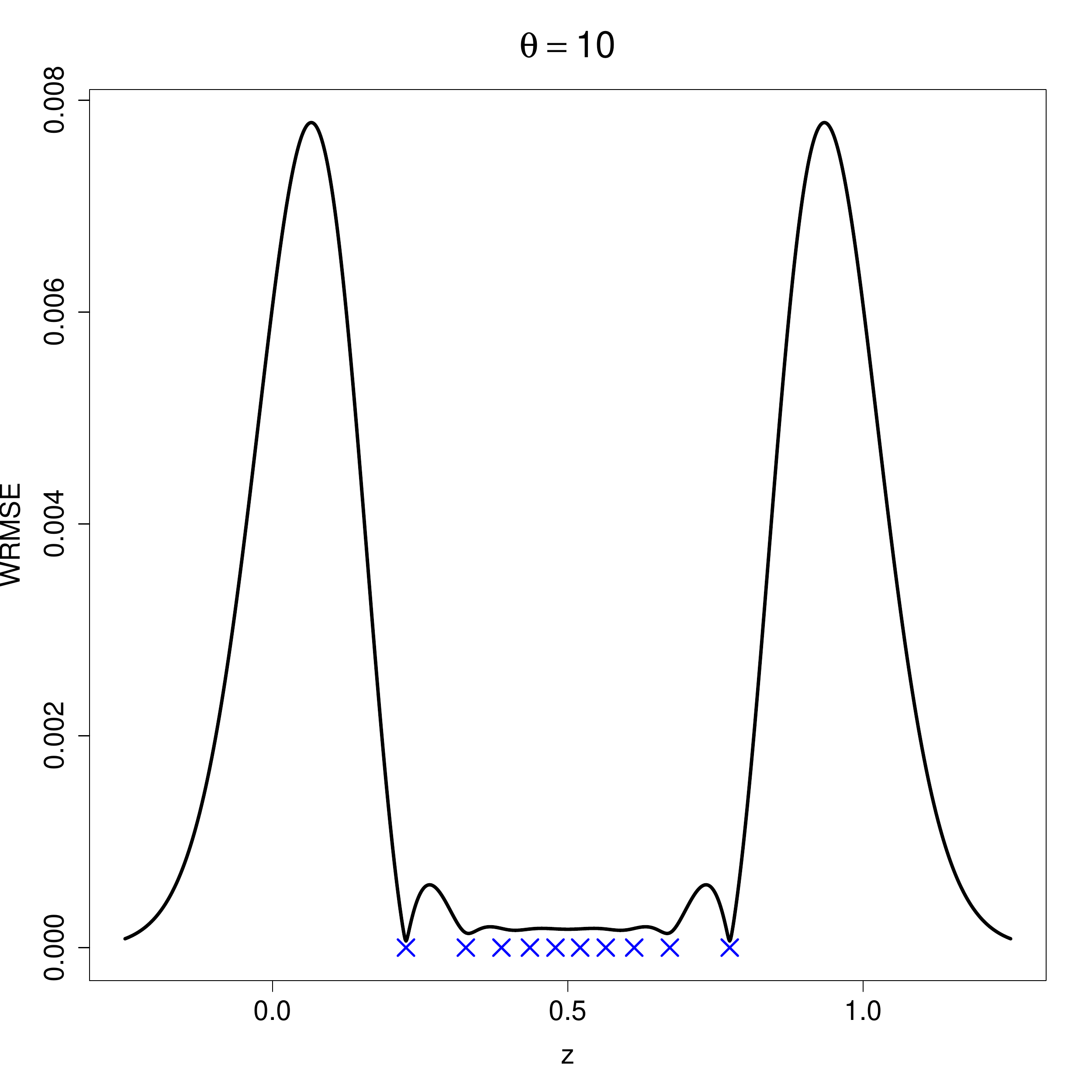} &
\includegraphics[width = 0.45\textwidth]{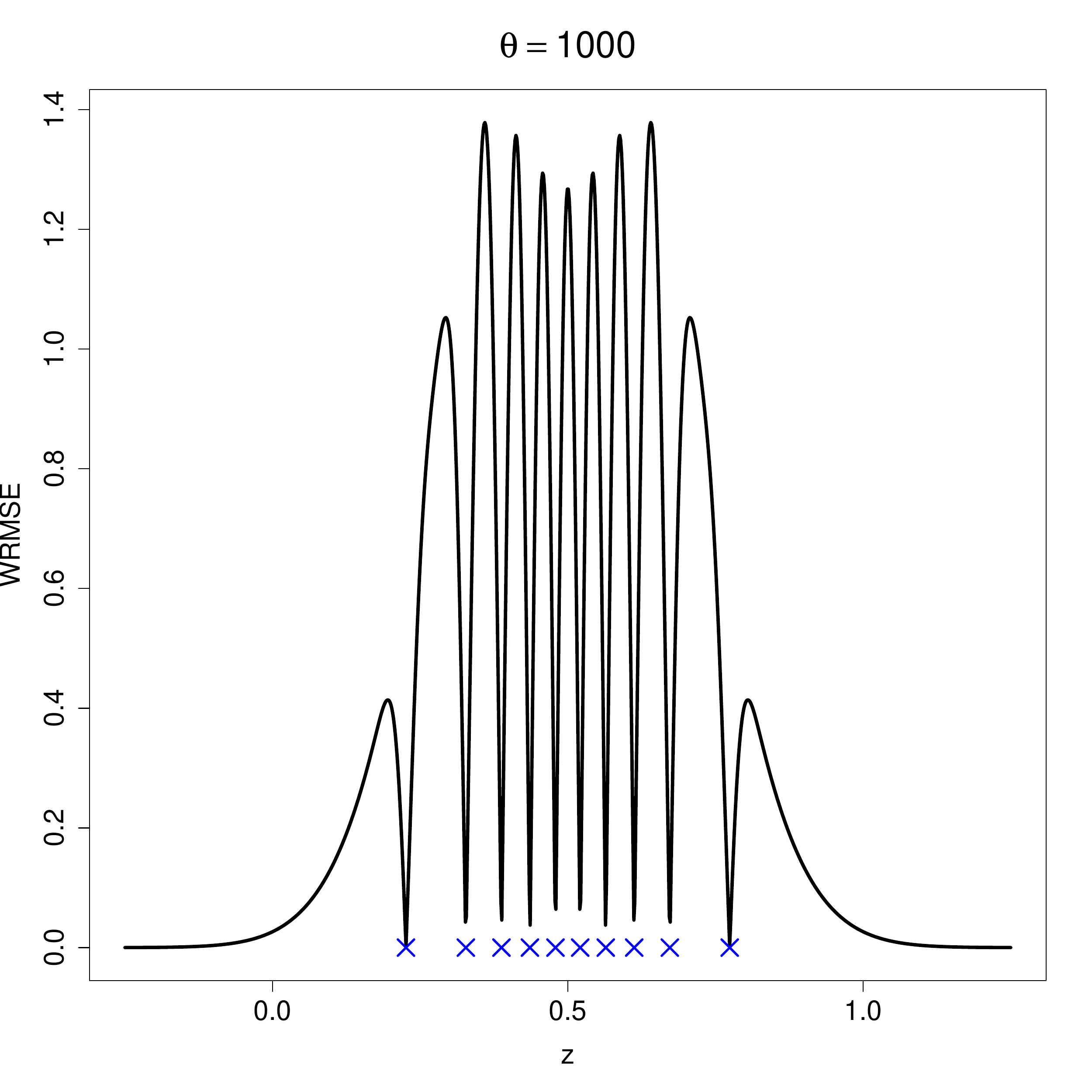}
\end{tabular}
\caption{Plot of weighted root mean squared error against the noise factor for a 10-point transformed design using $N(.5,\sigma)$ for $\theta=10$ (left) and $\theta=1000$ (right).}
\label{fig:TD}
\end{center}
\end{figure}

For illustration, let $z$ be a normal distribution with mean $0.5$ and standard deviation $\sigma=1/6$. These values are chosen so that the $\pm 3\sigma$ limits of $z$  coincide with $[0,1]$. Suppose we use $n=10$ and a Gaussian correlation function: $R(z_i-z_j)=\exp\{-\theta (z_i-z_j)^2\}$. As shown in \cite[p.19]{Fang1994}, $\bm d_0=\{.5/n, 1.5/n,\ldots,(n-.5)/n\}$ minimizes the discrepancy from a uniform distribution. So the desired design can be obtained as $F^{-1}(\bm d_0)$. Figure \ref{fig:TD} plots the weighted root mean squared error
\begin{equation}
WRMSE(\bm x,\bm z)=\sqrt{MSE(\bm x,\bm z)}f(\bm z),
\end{equation}
for $\theta=10$ and $\theta=1000$. The design points are shown as crosses in the same plots. We can see that the points are pulled towards the center as one would expect for a normal distribution. However, although the transformation seems to balance the WRMSE well throughout the region when $\theta=1000$, it seems to be too high in the tail regions when $\theta=10$. Thus, the benefit of using a transformed design seems to depend on the smoothness of the underlying response function. If the function is wiggly, then transformation will work great, but if the function is smooth, the transformation may do more harm than good. From our experience, the external noise factors usually have  a smooth relationship with the response. This could be because the realistic ranges of noise factors are much smaller than the possible ranges of control factors and therefore, the noise-response relationship can be adequately modeled using a smooth Gaussian process. Thus, $\theta$ is expected to be small in the Gaussian correlation function. In summary, a transformed design does not seem to be a good choice for the noise array and we need to look for other alternatives.

\graphicspath{{./Figures/}}
\subsection{Hybrid Design}
We can find the design that minimizes IRMSE in (\ref{eq:IRMSE}) for a given value of $\bm \theta$ or even better by maximizing the efficiency in (\ref{eq:RIRMSE}). However, as mentioned before, this optimization is hard to perform, especially in high dimensions. We propose a simple idea to overcome this problem. We will find the  optimal design for one factor, which is easier. The optimal design can be viewed as a transformation of a uniform design. Now we use this  optimal transformation on each column of the noise factors of a space-filling design. Since the final design is obtained using a combination of space-filling  and model-based optimal design criteria, we will call the design a \emph{hybrid design}.

To fix the idea, consider a single noise factor $z$. Let $\bm d^*=\{z_1^*,\ldots,z_n^*\}$ be the optimal design obtained using the model-based criterion
\[\max_{\bm d} \min_{\theta \in \Theta}\frac{IRMSE(\bm d^*(\theta),\theta) }{IRMSE(\bm d, \theta)},\]
where
\[\bm d^*(\theta)=arg\min_{\bm d} \int_\mathcal{Z} \sqrt{MSE(z; \bm d, \theta)} f(z) dz.\]
We want to emphasize that these optimizations are computationally much simpler than those of (\ref{eq:RIRMSE}) and (\ref{eq:IRMSE}) which use the full factor space.
Let $\{u_i^*=(i-.5)/n, i=1,\ldots,n\}$ be the uniform design points in $[0,1]$. It is easy to show that the optimal design points are distinct, that is $z_i^*\ne z_j^*$ for $i \ne j$. Therefore, there exists a one-to-one transformation:
\begin{equation}\label{eq:transform}
z_i^*=T(u_i^*), \;\textrm{for}\; i=1,\ldots,n.
\end{equation}
For the case of multiple factors, let $\bm U_z$ be a space-filling design with levels $\{u_1^*,\ldots,u_n^*\}$ for each of the noise factor. Then, the hybrid design can be obtained as
\begin{equation}\label{eq:hybrid}
\bm D^*_z=T(\bm U_z),
\end{equation}
where the transformation $T(\cdot)$ is applied on each element of $\bm U_z$.

\begin{figure}
\begin{center}
\includegraphics[width = 0.45\textwidth]{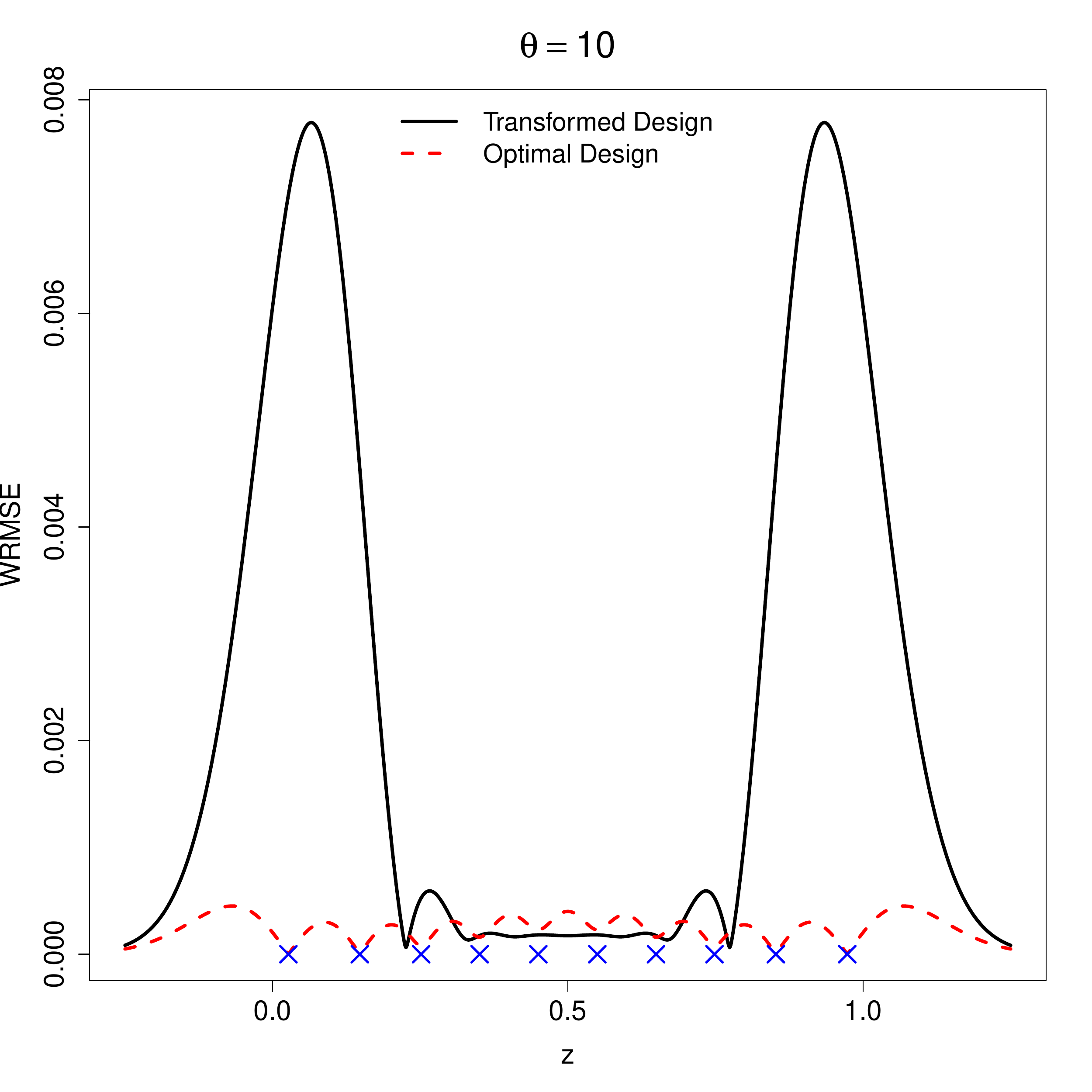}
\caption{Plot of weighted mean squared error against the noise factor for a 10-point transformed design and optimal design with $\theta=10$. The optimal design points are shown as crosses.}
\label{fig:OD}
\end{center}
\end{figure}

Consider again the same example of the previous subsection with $z\sim N(.5,1/6)$ and $\theta=10$. The WRMSE for the 10-run optimal design is plotted in Figure \ref{fig:OD}. We can see that the optimal design performs much better than the  transformed design. However, this improvement is not realizable in practice because we never know the true value of $\theta$ before the experiment. We need to choose a robust value of $\theta$ using (\ref{eq:RIRMSE}).

\graphicspath{{./Figures/}}

\begin{figure}
\begin{center}
\includegraphics[width = 0.45\textwidth]{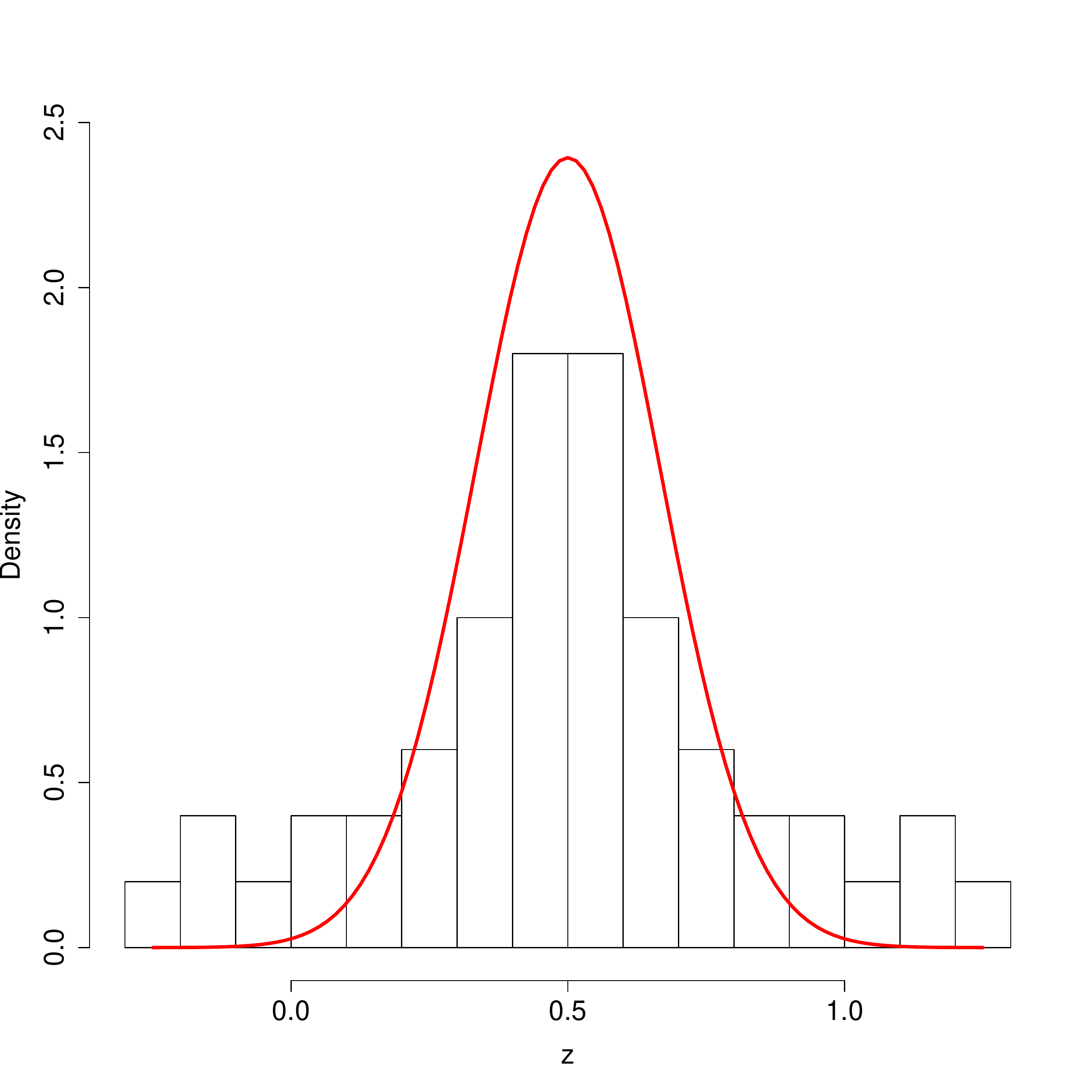}
\caption{Histogram of the IRMSE-optimal design obtained for $n=50$ and $\theta=30$. The density of the noise distribution is also shown.}
\label{fig:optdesign}
\end{center}
\end{figure}

\begin{figure}
\begin{center}
\includegraphics[width = 0.45\textwidth]{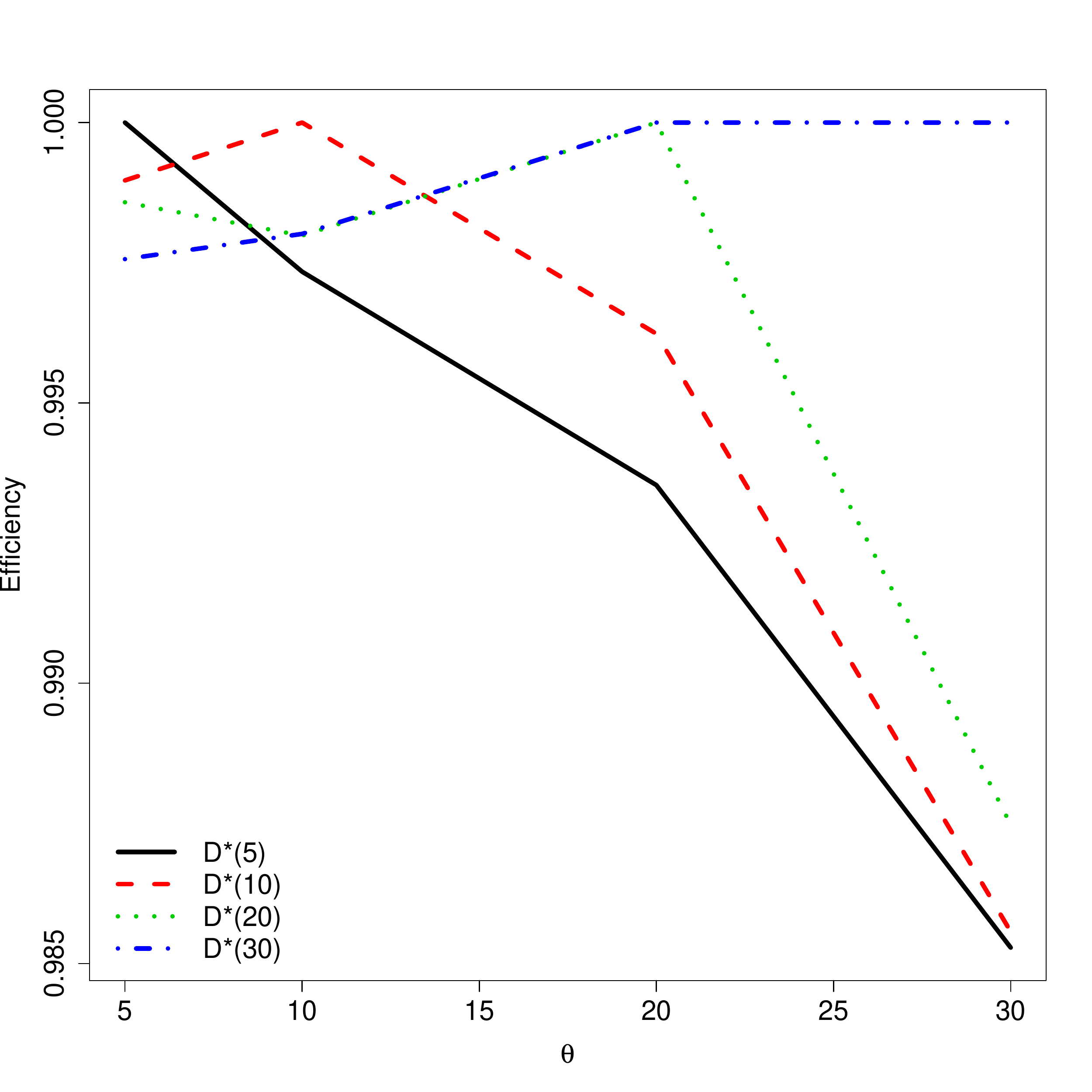}
\caption{Efficiency of the optimal design obtained for a given value of $\theta$ plotted against different values of $\theta$.}
\label{fig:efficiency}
\end{center}
\end{figure}

Consider a set $\Theta=\{5,10,20,30\}$ with $z\sim N(.5,1/6)$ and $n=50$. We first find the optimal designs for each of the four values of $\theta\in \Theta$. A histogram of the points for one case is shown in Figure \ref{fig:optdesign}. We can see that the IRMSE-optimal design points are slightly more dispersed than the noise distribution. The efficiencies of the four designs are computed using (\ref{eq:RIRMSE}) and are plotted over $\theta$ in Figure \ref{fig:efficiency}. We can see that the optimal design found based on smaller values of $\theta$ perform poorly for larger values of $\theta$. On the other hand, the optimal designs found using larger values of $\theta$ perform not so poorly for smaller values of $\theta$. This suggests that we should find the set of probable values of $\theta$ and use the largest value in that set to generate the optimal design.

\graphicspath{{./Figures/}}

\graphicspath{{./Figures/}}

Guessing the largest possible value of $\theta$ can be challenging in a practical problem. Moreover, the numerical inaccuracies in computing IRMSE increases with $n$ and the optimization becomes harder and unstable. Due to these difficulties, in the next subsection, we will try to identify an approximate optimal design that is easier to use in practical applications.

\subsection{Double Transformed Design}
Let $\bm d^*=\{z_1^*,\ldots,z_n^*\}$ be the optimal set of $n$ points that minimizes the IRMSE in (\ref{eq:IRMSE}) for a given value of $\theta$. The empirical distribution function of this point set is given by $F_n(z)=1/n\sum_{i=1}^n I(z_i^*<z)$, where $I(\cdot)$ is the indicator function. Our aim is to understand the limiting distribution of $F_n(z)$ as $n\rightarrow \infty$, which we denote by $\tilde{F}(z)$. This is not an easy problem because there is no explicit solution for the optimal design. Moreover, IRMSE is a complex function of the design points. Therefore, we will make use of an existing result on optimal designs for uniform distributions to get an idea of the limiting distribution.

For uniformly distributed variables, \cite{Dette2010} showed that a beta distribution with density $b(z; \alpha,\alpha)$ for $\alpha\in [0.5,1]$ is optimal for a reciprocal distance criterion which can be viewed as a surrogate for (\ref{eq:IRMSE}). See also \cite{Zhigljavsky2010} for a rigorous justification of this result. We know that if $F(\cdot)$ is the distribution function of $z$, then $F(z)\sim U(0,1)$. Thus, by using change of variables, the optimal density of the design can be obtained as
\begin{eqnarray}
\tilde{f}(z)&=&b(F(z); \alpha,\alpha)f(z)\nonumber\\
&=&\frac{\Gamma(2\alpha)}{\Gamma^2(\alpha)}\frac{f(z)}{\{F(z)[1-F(z)]\}^{1-\alpha}}.\label{eq:optdensityalpha}
\end{eqnarray}
Let $B_{\alpha}(z)=\int_0^z b(u;\alpha,\alpha) du$ be the distribution function of the beta distribution. Then, the asymptotic distribution function of the optimal design is given by
\begin{equation}\label{eq:optdist}
\tilde{F}(z)=B_{\alpha}\left(F(z)\right).
\end{equation}
Thus, if $\bm d_0$ denotes the uniform design, then an approximation to the optimal design can be obtained as
\begin{equation}\label{eq:optdesign}
\bm d^*=F^{-1}\left(B_{\alpha}^{-1}(\bm d_0)\right).
\end{equation}
To distinguish from the previous transformed design $F^{-1}(\bm d_0)$, we will call this the \emph{double transformed design}.

What should be the value of $\alpha$? If $\theta$ is large in the Gaussian correlation function, we should use $\alpha=1$, which leads to the transformed design. We have seen in Figure \ref{fig:TD} that the transformed design indeed works well with large $\theta$. However, as mentioned earlier, we are more interested in small values of $\theta$. \cite{Dette2010} recommended using $\alpha=1/2$, which is the limiting distribution of a $D$-optimal design for large degree polynomial regression. Theorem 10.1 in \cite{Fasshauer2016} shows that as the correlation parameter $\theta\rightarrow 0$, the Gaussian process predictor tends to a high degree polynomial interpolator and therefore, it makes sense to use an $\alpha$ value close to 1/2. The design points, in this case, are the same as Chebyshev nodes, which possess minimax optimality properties for polynomial interpolation \citep{Trefethen2013}. However, as discussed towards the end of previous subsection, the optimal design based on a small value of $\theta$ may work poorly for large values of $\theta$. Thus, intuitively, a value of $\alpha$ slightly larger than 1/2 such as 2/3 or 3/4 might be a more robust choice. We investigate this more carefully below for the case of a normal distribution.

It is easy to show that $MSE(z;\bm d)=1-\bm r(z)'\bm R^{-1}\bm r(z)\le 1-R^2(z-Q(z,\bm d))$, where $Q(z,\bm d)$ is the closest point in $\bm d$ to $z$. Let
\[\overline{IRMSE}(\bm d)=\int_{\mathcal{Z}}\sqrt{1-R^2(z-Q(z,\bm d))}f(z)dz,\]
which is an upper bound of $IRMSE(\bm d)$. For Gaussian correlation function,
\begin{eqnarray}
\overline{IRMSE}(\bm d)&=&\int_{\mathcal{Z}}\sqrt{1-\exp\{-2\theta(z-Q(z,\bm d))^2\}}f(z)dz \nonumber\\
&\approx &\sqrt{2\theta} \int_{\mathcal{Z}}|z-Q(z,\bm d)|f(z)dz,\label{eq:Zador}
\end{eqnarray}
where the approximation is valid for large $n$. \cite{Zador1982} has shown that the design that minimizes $\int |z-Q(z,\bm d)|^kf(z)dz$ has an asymptotic distribution proportional to $f^{1/(1+k)}(z)$. This implies that the asymptotic distribution of $\bm d$ that minimizes $\overline{IRMSE}(\bm d)$ should be proportional to $\sqrt{f(z)}$. Now consider the case of a normal distribution $f(z)=\phi(z;.5,\sigma)$. Based on the simple approximation given by \cite{Bell2015}, $F(z)[1-F(z)]\approx f^{4/\pi}(z)$. Substituting this approximation in (\ref{eq:optdensityalpha}), we obtain $\tilde{f}(z)\approx \Gamma(2\alpha)/\Gamma^2(\alpha) \{f(z)\}^{1-4(1-\alpha)/\pi}$. This will be proportional to $\sqrt{f(z)}$ if $\alpha=1-\pi/8\approx 0.607$. Similar exercise using IMSE gives $\alpha=1-\pi/12\approx 0.738$. Based on these values, we choose $\alpha=2/3 \approx (.607+.738)/2$. Interestingly, in a totally different problem setting of searching for the maximum of a continuous function using non-adaptive algorithms, \cite{Al1996} showed that the same beta density with $\alpha=2/3$ is optimal.

\graphicspath{{./Figures/}}
\begin{figure}
\begin{center}
\includegraphics[width = 0.45\textwidth]{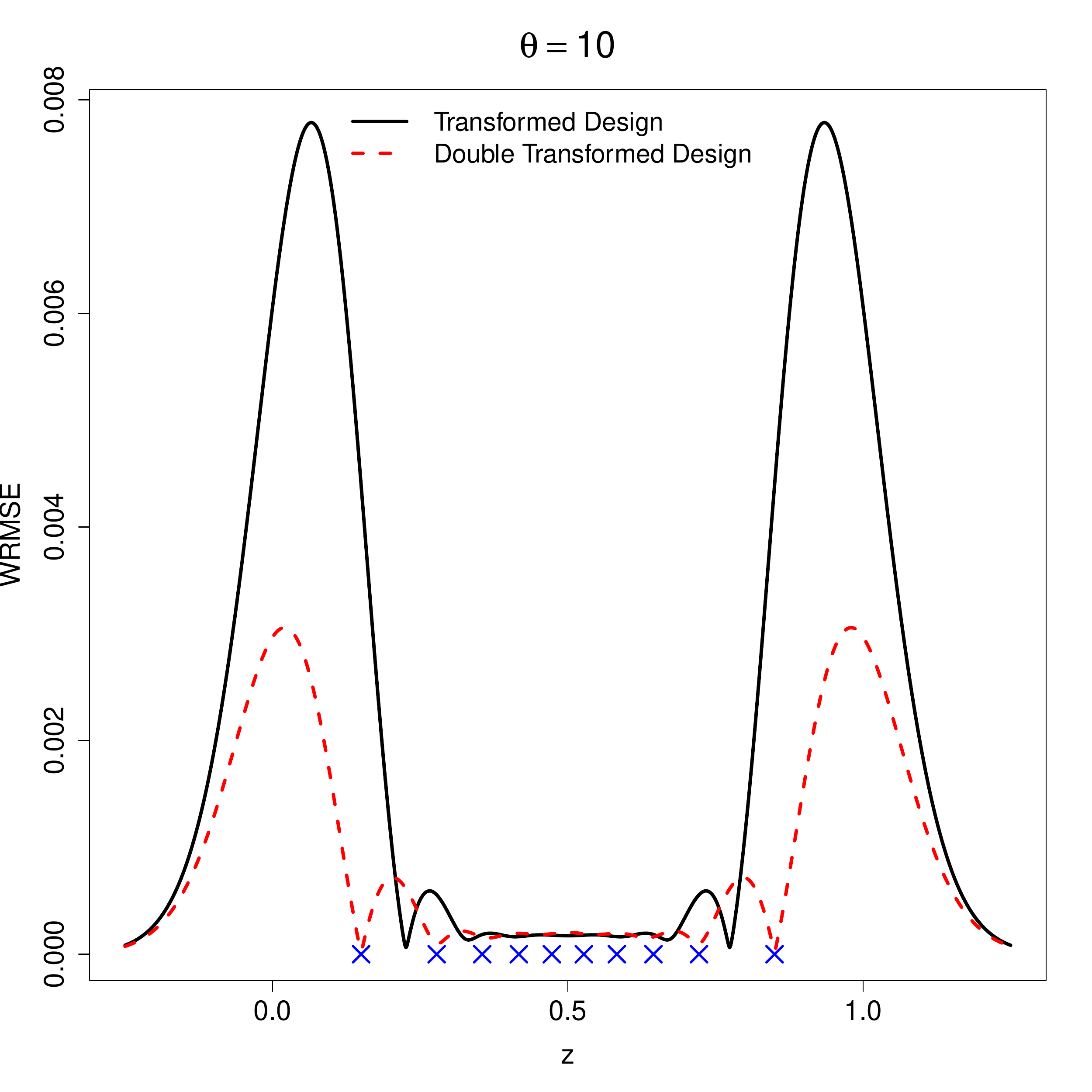}
\caption{Plot of weighted mean squared error against the noise factor for a 10-point transformed design using $N(.5,\sigma)$, and double transformed design using (\ref{eq:optdesign}) with $\alpha=2/3$ for $\theta=10$.}
\label{fig:DT10}
\end{center}
\end{figure}
Consider the previous example with $z\sim N(.5,1/6)$.  The WRMSE for the double transformed design using (\ref{eq:optdesign}) with $\alpha=2/3$ is shown in Figure \ref{fig:DT10}. We can see that WRMSE for the double transformed design is much smaller than that using the transformed design, but not as good as the optimal design in Figure \ref{fig:OD}.

\begin{figure}[h]
\begin{center}
\begin{tabular}{cc}
\includegraphics[width = 0.45\textwidth]{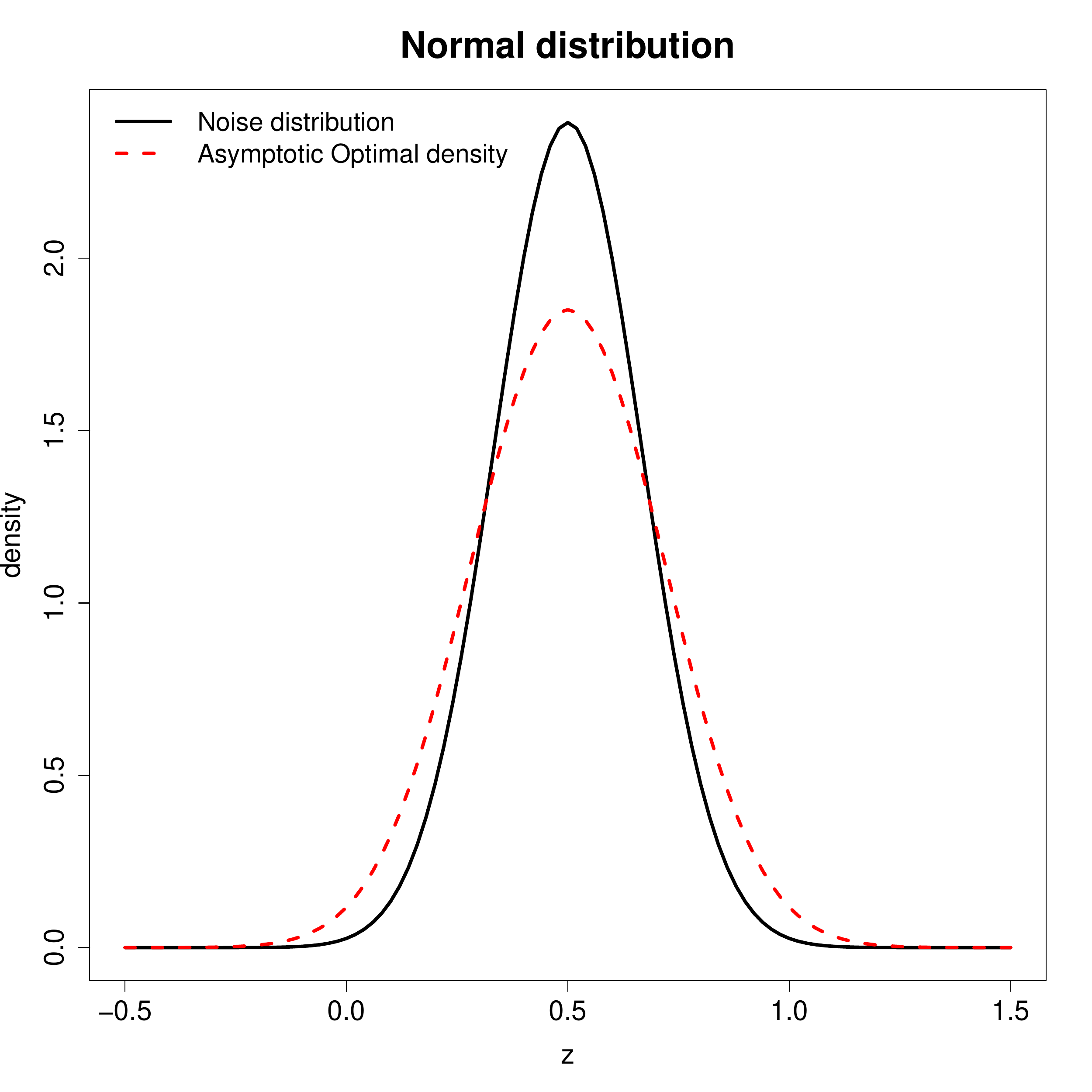} &
\includegraphics[width = 0.45\textwidth]{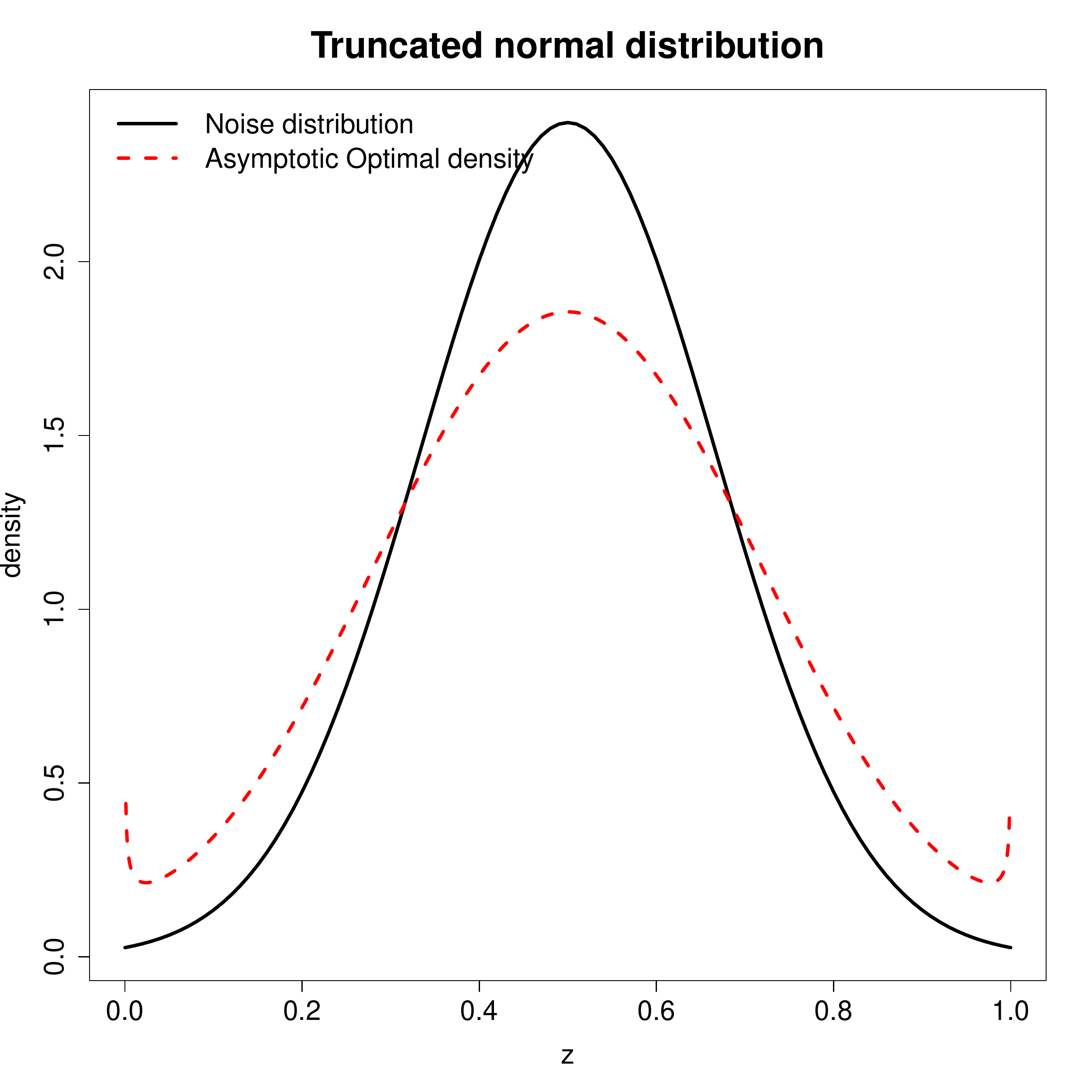}
\end{tabular}
\caption{Comparison of $N(.5,\sigma)$ (left) and the asymptotic optimal density in (\ref{eq:optdensityalpha}) with $\alpha=2/3$. The right panel shows the densities when the normal distribution is truncated in $[0,1]$.}
\label{fig:optdensity}
\end{center}
\end{figure}

The left panel of Figure \ref{fig:optdensity} compares the noise distribution with the asymptotic optimal density in (\ref{eq:optdensityalpha}). We can see that the density for the optimal design is more dispersed than the original noise distribution.  The right panel of Figure \ref{fig:optdensity} shows the optimal density when the noise distribution is truncated to $[0,1]$. This density has three modes: one at the center and two at the boundaries. This is an interesting result, because Taguchi (1987) has recommended using three levels for the noise factor, one at the mean and two at the extremes. The result in Figure \ref{fig:optdensity} can be viewed as an extension of this three-level design for physical experiment to an $n$-level design for computer experiment, where projections are important.

\graphicspath{{./Figures/}}

A similar investigation using the initial IMSE criterion in (\ref{eq:IMSE0}) shows that $\alpha\approx 0.476$ is the optimal choice. However, this makes the optimal density quite dispersed and places points in very low probability regions. For example, when $n=100$, the points can be as far as $\pm 3.95\sigma$ from the center. On the other hand, $\alpha=2/3$ places points within $\pm 3.25\sigma$ from the center, which looks more reasonable. This is why we feel the IRMSE criterion in (\ref{eq:IRMSE}) or the IMSE criterion in (\ref{eq:IMSE}) is more meaningful than the IMSE criterion in (\ref{eq:IMSE0}). This was also verified using the prediction performance on some test cases.

\section{Factors with Internal Noise}
There are some factors whose nominal values can be controlled, but they can vary around their nominal values. Such factors are said to have \emph{internal noise}. Examples include, part-to-part variability within their manufacturing tolerances and process parameter variability around its target. On the other hand, \emph{external noise} factors are completely uncontrollable including their nominal values. Examples of external noise factors include user conditions, incoming raw material properties,  etc. In this section we will propose methods for designing experiments with internal noise factors. One may wonder why we need to consider these factors differently from the external noise factors. Why not just merge them with the external noise factors and use the techniques described in the previous two sections? The reason is that we don't need to vary the internal noise factors in the experiment! They can be easily introduced at the modeling stage. Thus, internal noise factors can be ignored at the design stage although this may not be the ``optimal'' approach. This topic has received scant attention in the literature except possibly for the work of \citet{Kang2009} for the case of physical experiments.

A factor with internal noise can be represented as $X=x+e$, where the nominal value $x$ is controllable and the internal noise $e$ is uncontrollable. Here we have used additive noise, but the case of multiplicative noise can be handled similarly. Suppose we have estimated the relationship with the nominal values of the factors ($\bm x$) and external noise factors ($\bm z$): $\hat{g}(\bm x,\bm z)$, then we can easily obtain the relationship with the internal noise as $\hat{g}(\bm x+\bm e,\bm z)$. This is why we don't need to vary $\bm e$ in the experiment.

Since a factor with internal noise is both a ``control''  and a ``noise'' factor, it make sense to cross this factor with the other factors in the experiment. Thus, if $\bm D_X$ denotes the design for the factors with internal noise, then we can obtain the cross array using $\bm D_x \times \bm D_z \times \bm D_X$. From this we can obtain the jittered cross array using the same algorithm discussed earlier. Now, we only need to decide on how to choose the levels for a factor with internal noise.

As before let's assume $x$ to follow a uniform distribution in $[0,1]$ and $e$ to have a noise distribution with density $f_e(e)$. Because the internal noise factor is the result of not controlling the process well, it is mostly going to have a normal distribution.  So let $e\sim N(0,\sigma_e)$. The optimal design $\bm d^*=\{x_1^*,\ldots,x_n^*\}$ can be obtained by minimizing
\[IRMSE(\bm d,\theta)=\int_0^1\int_{-\infty}^{\infty}\sqrt{1-\bm r(x+e)'\bm R^{-1}\bm r(x+e)}\phi(e;0,\sigma_e)de dx,\]
where $\phi (e; 0,\sigma_e)$ denote the density of a normal distribution with mean 0 and standard deviation $\sigma_e$. Here it is better to consider the IMSE criterion in (\ref{eq:IMSE}) because an explicit expression for the integral can be obtained under a Gaussian correlation function, $R(h)=\exp(-\theta h^2)$. Thus,
\begin{eqnarray}
IMSE(\bm d,\theta)&=& \int_0^1\int_{-\infty}^{\infty}\{1-\bm r(x+e)'\bm R^{-1}\bm r(x+e)\}\phi^2(e;0,\sigma_e)/C_2de dx\\
&=& \frac{1}{2\sigma_e\sqrt{\pi}C_2}\left[1-\int_0^1 tr\left\{\bm R^{-1}\bm A(x)\right\}dx\right], \label{eq:internal}
\end{eqnarray}
where the $ij$th element of $\bm A(x)$ is given by
\[\bm A_{ij}(x)=\frac{1}{\sqrt{1+2\theta \sigma_e^2}}\exp\left\{-\frac{2\theta}{1+2\theta\sigma_e^2}\left(x-\frac{x_i+x_j}{2}\right)^2\right\}\exp\left\{-\frac{\theta}{2}(x_i-x_j)^2\right\},\]
for $i,j=1,\ldots,n$. The integration with respect to $x$ can also be done explicitly to obtain
\begin{equation}\label{eq:internal}
IMSE(\bm d,\theta)=\frac{1}{2\sigma_e\sqrt{\pi}C_2}\left[1- tr\left\{\bm R^{-1}\overline{\bm A}\right\} \right],
\end{equation}
where
\[\overline{\bm A}_{ij}=\frac{\sqrt{\pi}}{\sqrt{2\theta}}\left\{ \Phi\left( \frac{\sqrt{\theta}(2-x_i-x_j)}{\sqrt{1+2\theta\sigma^2_e}}\right)-\Phi\left( \frac{-\sqrt{\theta}(x_i+x_j)}{\sqrt{1+2\theta\sigma^2_e}}\right)\right\} \exp\left\{-\frac{\theta}{2}(x_i-x_j)^2\right\},\]
where $\Phi(\cdot)$ is the standard normal distribution function. Similar explicit expressions could have been obtained in the previous section as well for the external noise factors, but we didn't do it because of numerical issues. We found the formula in (\ref{eq:internal}) to be quite vulnerable to numerical issue when $\bm R$ is nearly singular, which happens when $\theta$ is small. For the external noise factors, we need to consider small values of $\theta$ because they are expected to have a smooth relationship with $y$. On the other hand, the relationship with $x$ can be quite rough and therefore, here we should use large values of $\theta$ which doesn't lead to numerical problems.

\begin{figure}
\begin{center}
\includegraphics[width = 0.45\textwidth]{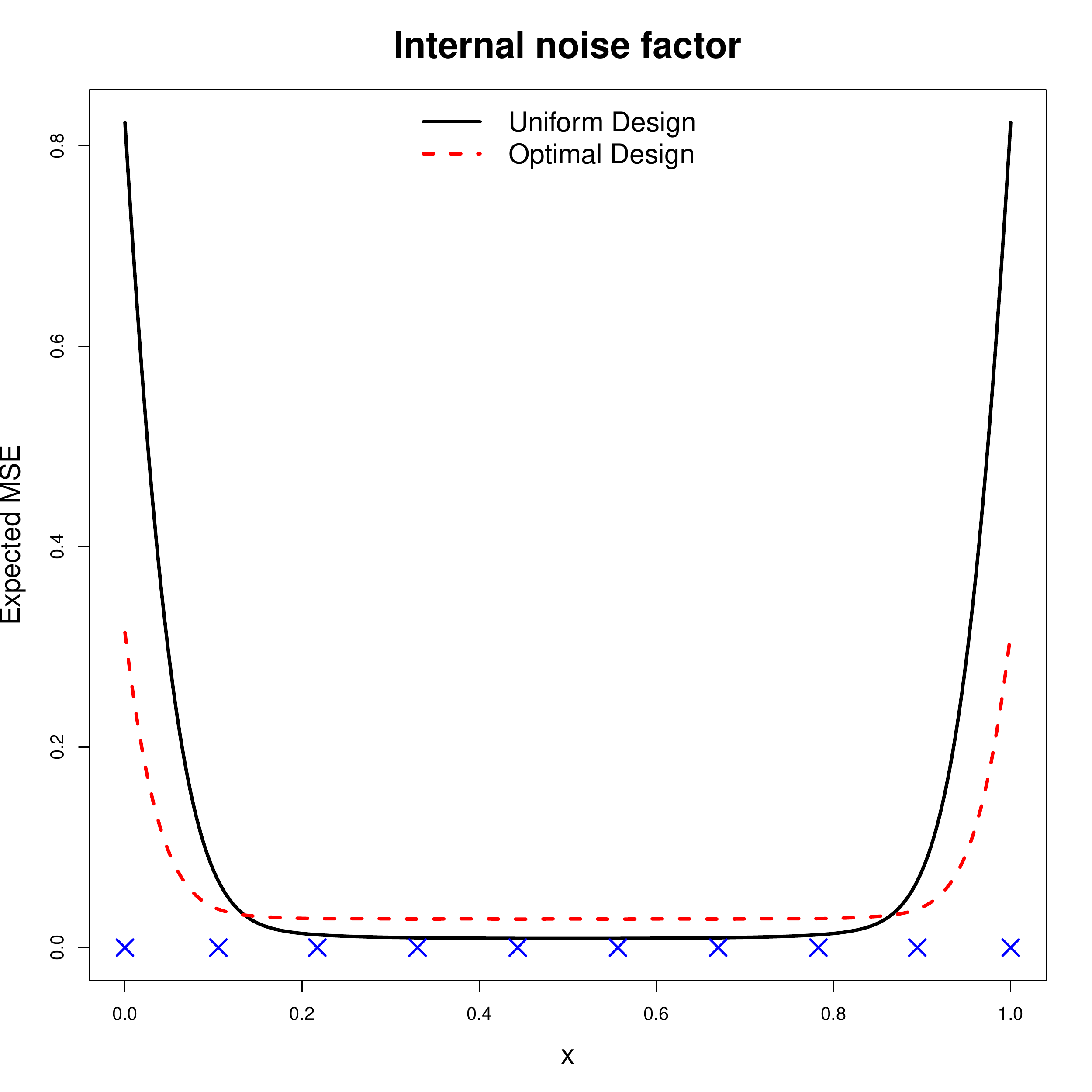}
\caption{Comparison of expected MSE for the 10-point uniform design (solid) and optimal design (dashed) when $\sigma_e=1/12$ and $\theta=50$. The optimal design points are shown as crosses.}
\label{fig:internal}
\end{center}
\end{figure}

As an example, consider a factor with internal noise distribution $e\sim N(0,\sigma_e)$. Let $\sigma_e=1/12$ and $\theta=50$ in the Gaussian correlation function. A 10-point optimal design is obtained by numerically minimizing (\ref{eq:internal}). Figure \ref{fig:internal} plots the expected mean squared error for the 10-point uniform design and the optimal design. We can see that the optimal design is almost equally spaced but  with points placed at the boundaries. Our simulations show that $\{0,1/(n-1),2/(n-1),\ldots,1\}$ is close to optimal, which is only a slight change from the uniform design.

\graphicspath{{./Figures/}}

\section{EXAMPLES}
\subsection{A Simulated Example}
\begin{figure}[h]
\begin{center}
\includegraphics[width = 0.75\textwidth]{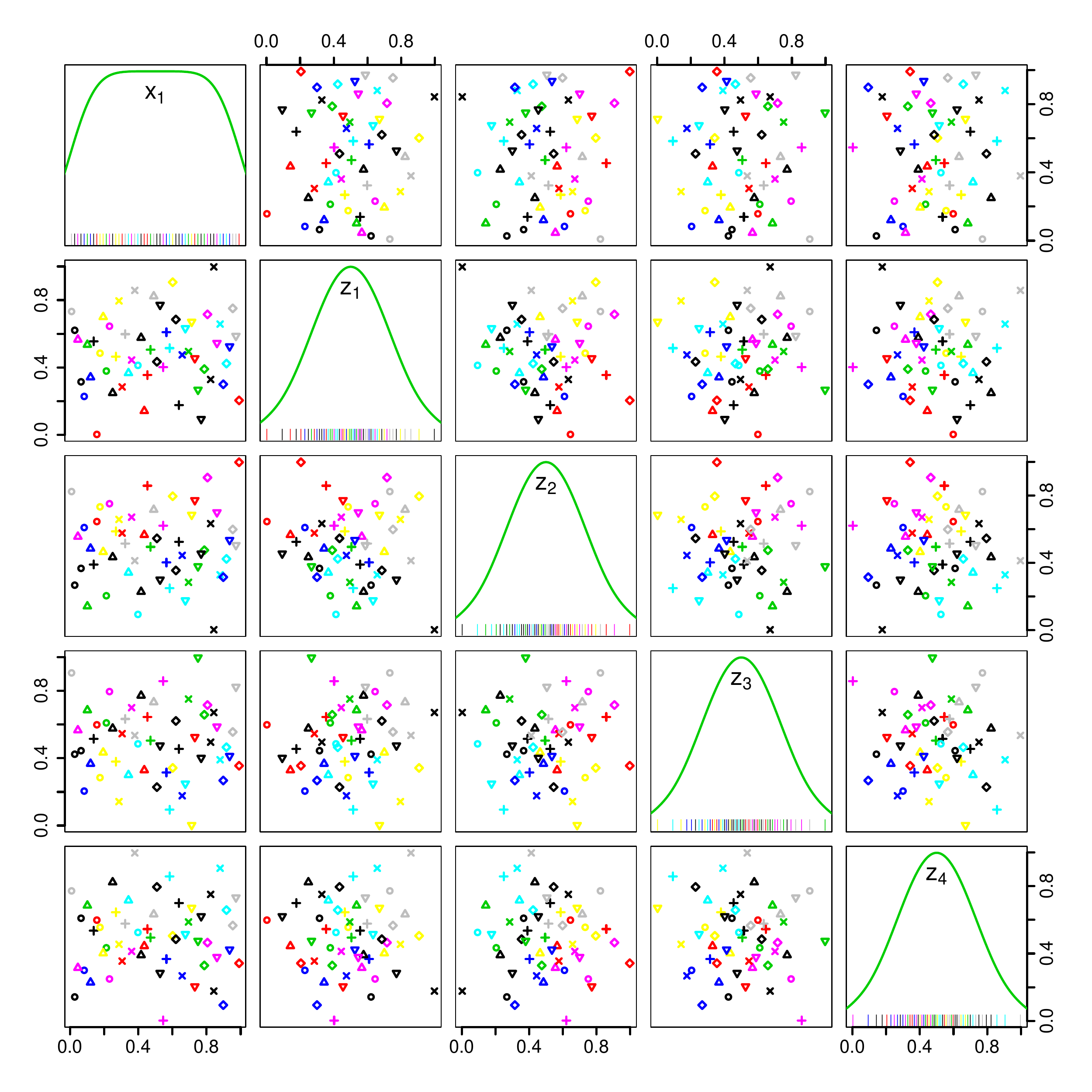}
\caption{Two-dimensional projections of double transformed jittered cross array (DTJCA) in the simulation example.}
\label{fig:DTJCA}
\end{center}
\end{figure}

\begin{figure}[h]
\begin{center}
\begin{tabular}{cc}
\includegraphics[width = 0.45\textwidth]{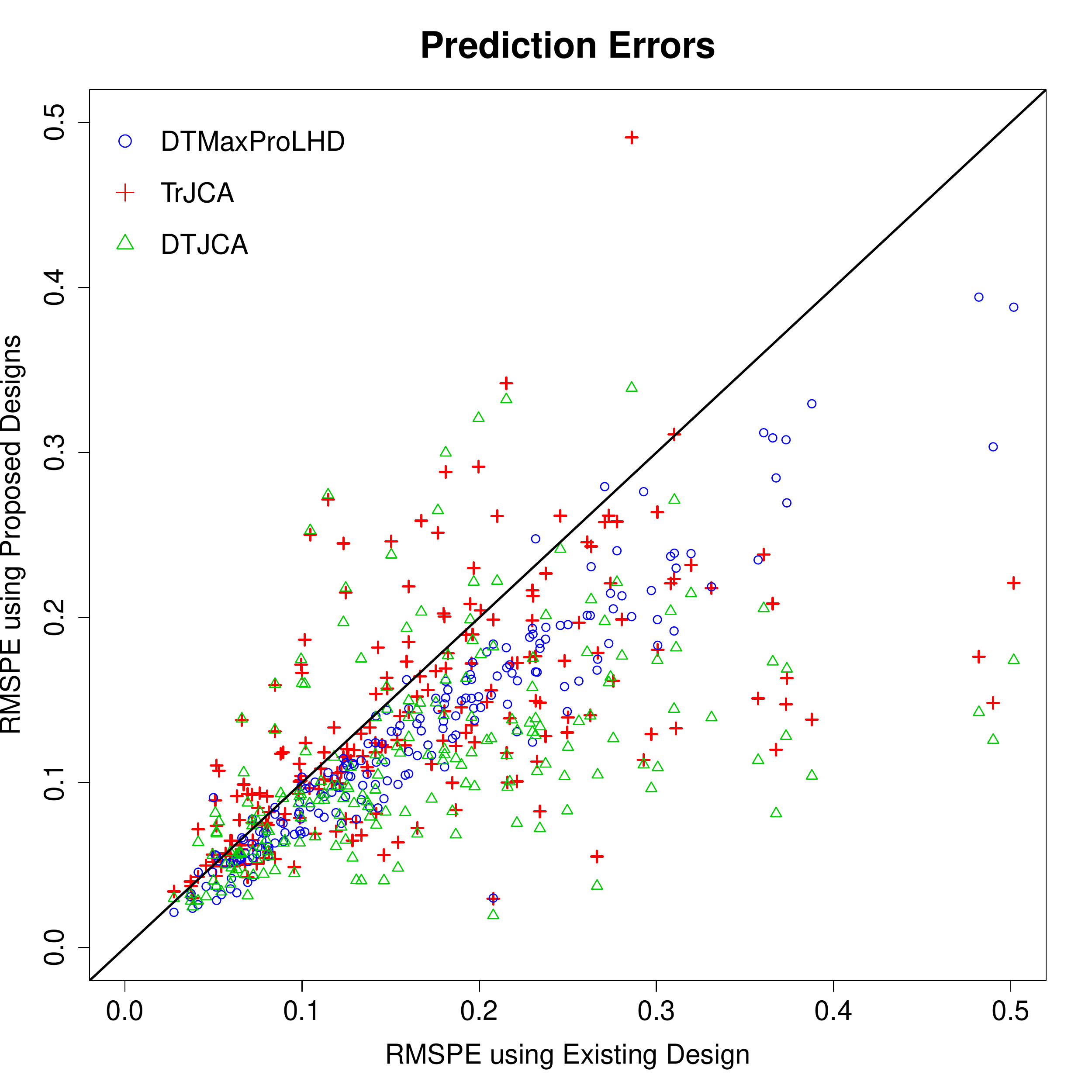} &
\includegraphics[width = 0.45\textwidth]{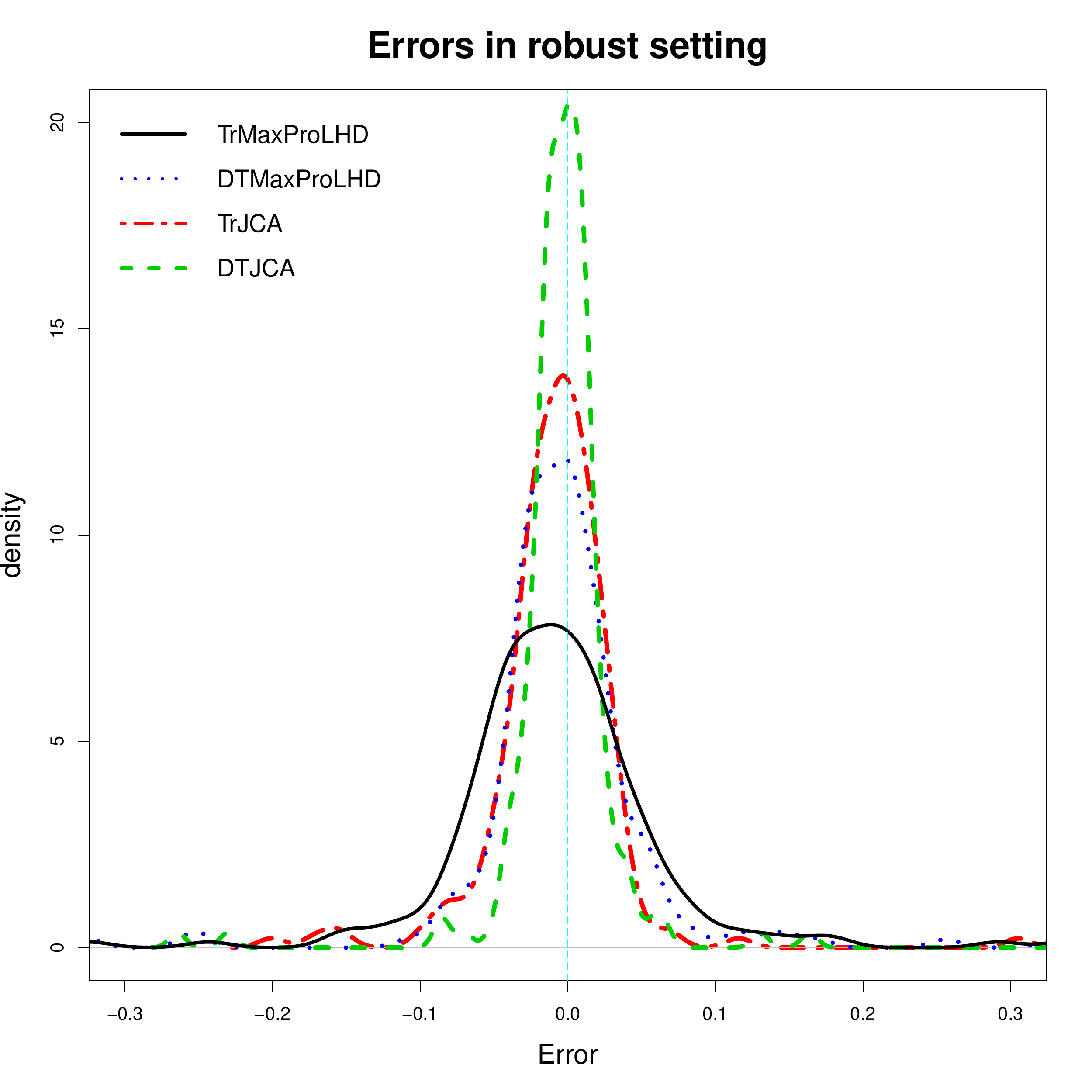}
\end{tabular}
\caption{Comparison of RMSPE (left) of kriging models fitted using the data generated by the existing transformed MaxProLHD (x-axis) and proposed designs (y-axis) (double transformed MaxProLHD, transformed JCA, and double transformed JCA) in the simulated example. The right panel shows the density plots of errors in the robust settings based on the four designs.}
\label{fig:toyexample}
\end{center}
\end{figure}

Consider a simple example with one control and four external noise factors. Let
\[y=\sum_{i=1}^4\beta_i(x-\gamma_i)z_i^2e^{-(x-\gamma_5)^2},\]
where $z_1,z_2\sim^{iid} N(0,1)$. We constructed a cross array using $n_1=6$ equally spaced levels for the control factor and an MmLHD with $n_2=9$ runs for the noise factors. From this a JCA is obtained using the sequential MaxPro algorithm described in Section 3. For comparison, we also constructed a MaxProLHD in 54 runs. We consider two versions of these two designs using transformed noise array and double transformed noise array. The resulting four designs are denoted as TrMaxProLHD, DTMaxProLHD, TrJCA, and DTJCA. Two-dimensional projections of DTJCA is shown in Figure \ref{fig:DTJCA}. We simulated 200 cases by randomly sampling $\beta_i$'s and $\gamma_i$'s from $U(0,1)$ and kriging models were fitted using the data generated by each of the three designs. 100 test samples were generated using a scrambled Sobol sequence with the noise factor columns transformed using the inverse distribution function of the noise distribution. The root-mean squared prediction errors from the kriging models are plotted in the left panel of Figure \ref{fig:toyexample}. We can see that the the double transformation on the MaxproLHD and JCA has helped to improve the prediction errors.

\graphicspath{{./Figures/}}

\graphicspath{{./Figures/}}

Suppose our aim is to minimize the variance of the response due to the noise factors. We have computed the true robust setting of the control factor ($x^*$) by minimizing the variance and also the robust settings obtained from the fitted models based on the four designs ($x_1^*,x_2^*,x_3^*$, and $x_4^*$). The right panel of  Figure \ref{fig:toyexample} shows the density plots of the errors $x_i^*-x^*$, for $i=1,2,3,4$. We can see that the DTJCA gives the best performance followed by TrJCA and DTMaxProLHD. Thus, although JCA didn't help improve the prediction errors over the existing TrMaxProLHD, it does seems to improve the identification of the robust setting.

\subsection{A Real Example}
\begin{figure}
\begin{center}
\includegraphics[width = 0.35\textwidth]{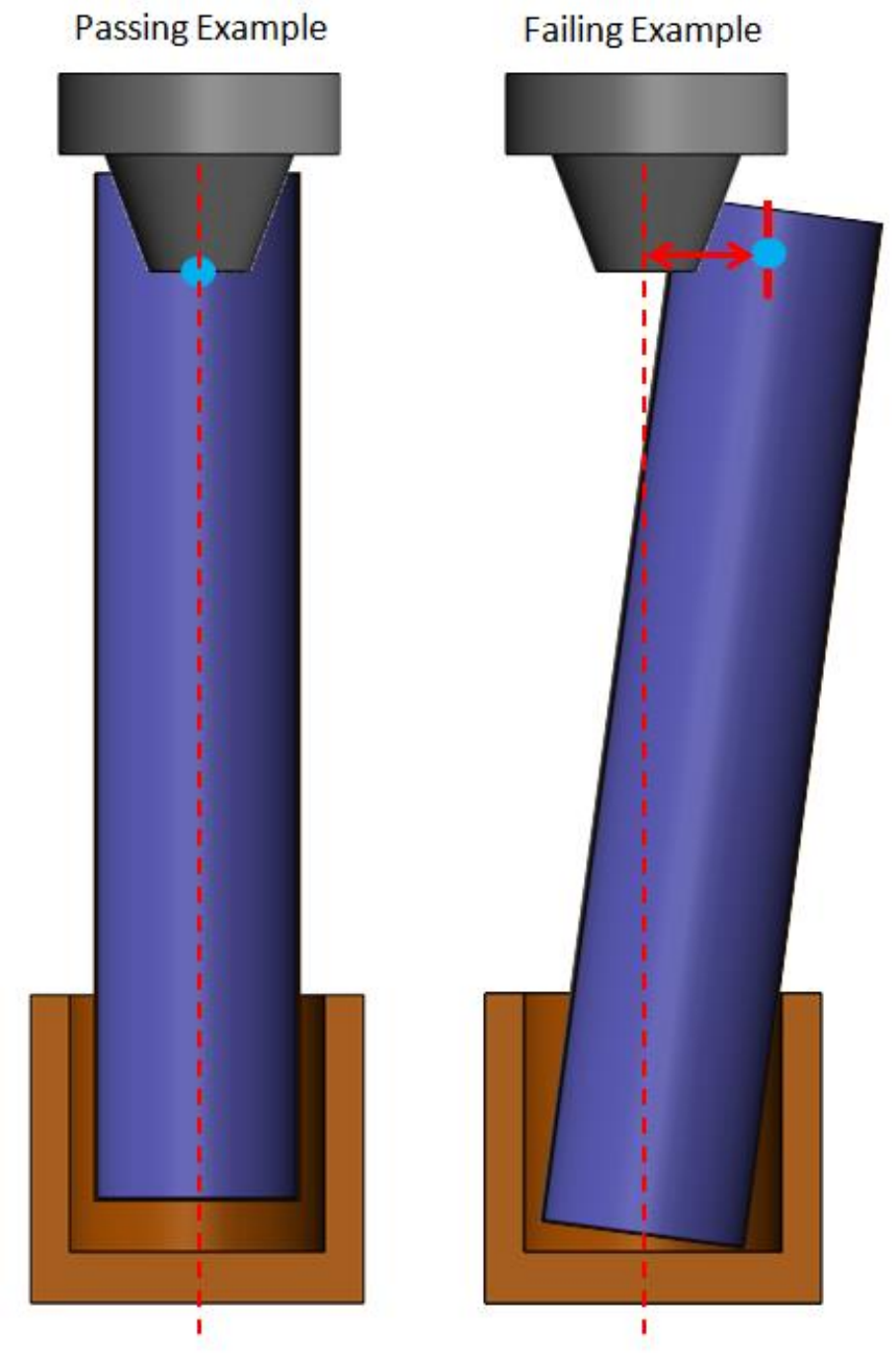}
\caption{Illustration of the package holder deflection. Large deflections can cause quality problems.}
\label{fig:realexample}
\end{center}
\end{figure}

Computer experiments with noise factors are quite common for the simulations conducted at The Procter \& Gamble Company. The specific example we will use in our study involves a manufacturing packing line.  The example has been slightly modified for the benefit of simplicity and to prevent disclosure of any potential sensitive information.  A computer simulator was developed for one critical transformation of the packing line involving transport of the package for product fill. A computer experiment with nine input factors was performed and an emulator was built. In this study, we use this emulator for investigating the robustness. Variables $x_1, x_2, x_3, x_4, x_5$, and  $x_6$ are process variables such as speed of the line and dimensions of the puck that transports the package that will remain fixed or easy to control once they are chosen and therefore are defined as control factors. Variables $z_1$, $z_2$, and $z_3$ are material properties  of the packaging component such as density and modulus which are defined as noise factors given that there is variation in normal production of the material supplier.  The output response from the computer simulation measures the deflection (deviation from a vertical orientation) after the package holder stops on the packing line, which impacts the quality of the given packing line transformation.  Figure \ref{fig:realexample} illustrates the ``passing'' and ``failing'' scenarios of the package holder. The objective of this study is to find the settings for the six control factors that are robust to the variation of the three noise factors. From historical data, the noise factors are found to be approximately normally distributed. After re-scaling, we let $z_i\sim^{iid} N(.5,\sigma)$ for $i=1,2,3$ with $\sigma=1/6$.

\graphicspath{{./Figures/}}

\begin{figure}[h]
\begin{center}
\begin{tabular}{cc}
\includegraphics[width = 0.45\textwidth]{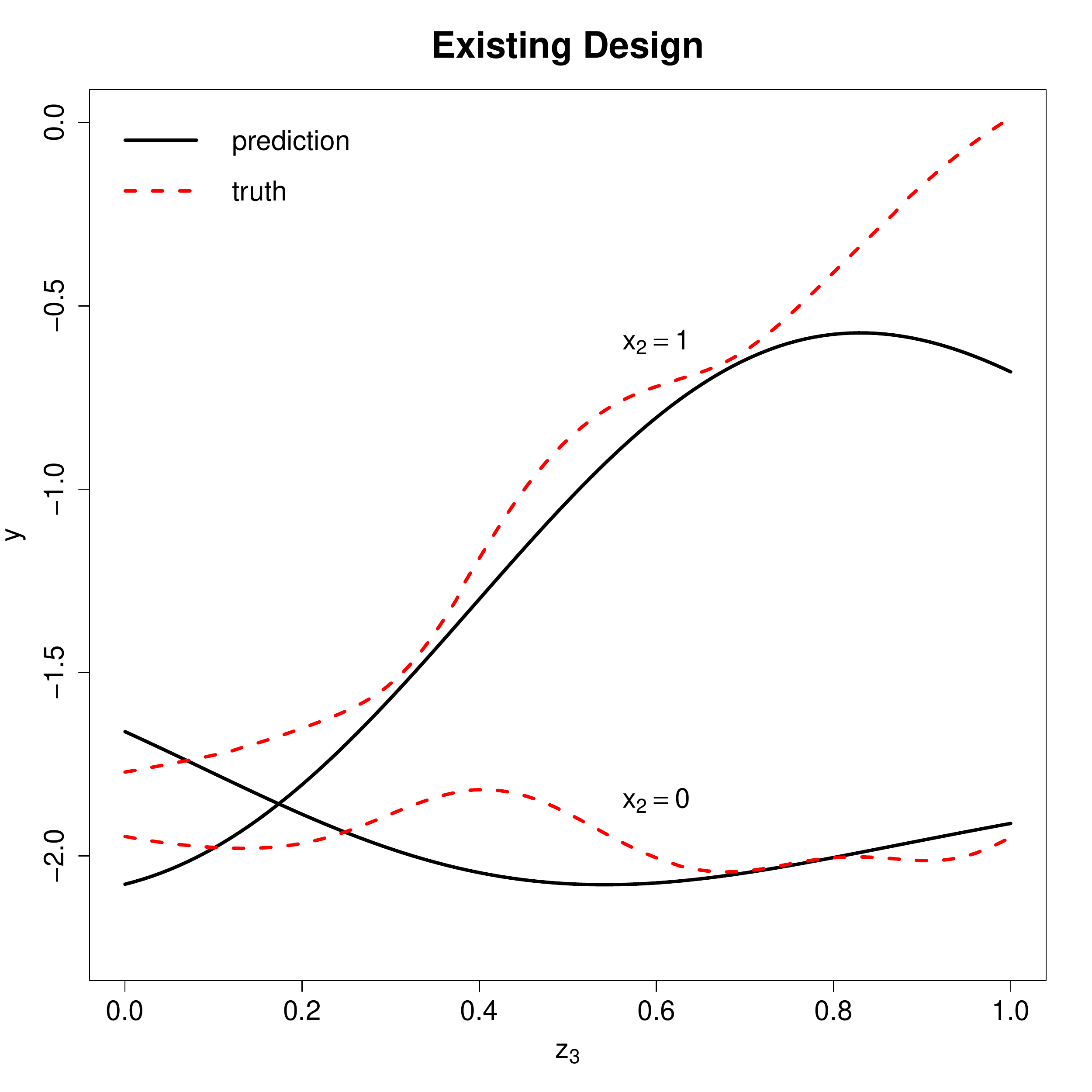} &
\includegraphics[width = 0.45\textwidth]{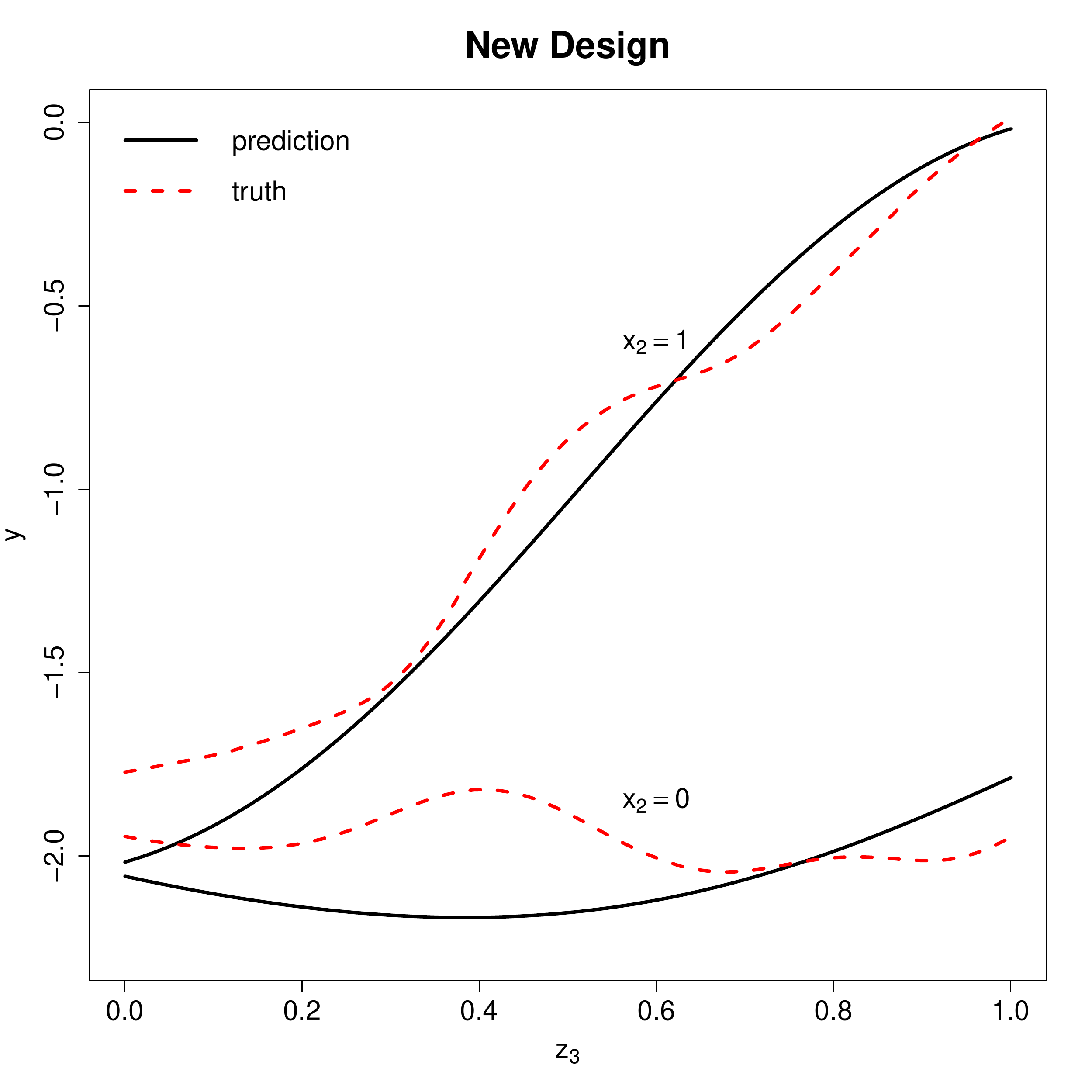}
\end{tabular}
\caption{Interaction plot of $x_2$ against $z_3$ in the real example. The left panel shows the interaction obtained using the existing transformed MaxProLHD and the right panel using the new double transformed JCA.}
\label{fig:realexampleresult}
\end{center}
\end{figure}

First we generated an MmLHD with $n_1=13$ runs for the six control factors and an MmLHD with $n_2=7$ runs for the noise factors. The JCA in 91 runs is thus obtained using the sequential MaxPro algorithm and then performed the double transformation on the noise factor columns using (\ref{eq:optdesign}). We also generated a MaxProLHD in 91 runs for comparison and transformed using the noise distribution. Kriging models were fitted to the data generated from the two designs. We found the prediction errors from the two fitted models to be close, but there was some major differences in the estimation of control-by-noise interactions. Figure \ref{fig:realexampleresult} shows the interaction between $x_2$ and $z_3$, which is the most significant interaction in this experiment. Clearly, the new double transformed JCA did a much better job in accurately estimating the interaction than the existing transformed MaxProLHD.

\graphicspath{{./Figures/}}

\section{CONCLUSIONS}
In this paper we have proposed space-filling designs that are suitable for identifying robust settings using computer experiments. The key idea was to modify the well-known cross array designs using a space-filling criterion such as the maximum projection criterion. We have also proposed how to optimally choose the noise array. The most intuitive way to construct the noise array is to transform the columns of the array using the inverse cumulative distribution function of the noise factors. This will pull the design points to the high probability region of the noise distribution. However, we found this ``pulling effect'' to be too extreme. This was mainly because the response is usually a smooth function of the noise factors and to precisely estimate a smooth function it is desirable to push the points outward from the center. The optimal design balances this ``pulling'' and ``pushing'' effects. We found that pushing the points uniformly distributed in the unit interval using a $Beta(2/3,2/3)$ distribution before applying the inverse probability transform to be close to asymptotically optimal for normally distributed noise variables. We have also proposed model-based methods to obtain the optimal transformation for any noise distribution, but it requires specification of the upper bound of certain correlation parameters.

\begin{center}
    {\Large\bf ACKNOWLEDGMENTS}
\end{center}
This research is supported by a U.S. National Science Foundation grant DMS-1712642 and a U.S. Army Research Office grant W911NF-17-1-0007.

\begin{center}
{\bf APPENDIX: Proof of Theorem 1}
\end{center}
Consider a production correlation $R(\bm x_i-\bm x_j,\bm z_i-\bm z_j)=R_x(\bm x_i-\bm x_j)R_z(\bm z_i-\bm z_j)$. Now, if $\bm D=\bm D_x \times \bm D_z$, then $\bm R=\bm R_x \otimes \bm R_z$ and $\bm r(\bm x,\bm z)=\bm r_x(\bm x) \otimes \bm r_z(\bm z)$, where $\otimes$ denotes Kronecker product (see, for example, Hung et al. 2015). Then, by using the properties of Kronecker products
\begin{eqnarray*}
\bm r(\bm x,\bm z)'\bm R^{-1}\bm r(\bm x,\bm z) &=& (\bm r_x(\bm x) \otimes \bm r_z(\bm z))'(\bm R_x \otimes \bm R_z)^{-1}\bm r_x(\bm x) \otimes \bm r_z(\bm z)\\
&=&(\bm r_x(\bm x)' \otimes \bm r_z(\bm z)')(\bm R_x^{-1} \otimes \bm R_z^{-1})\bm r_x(\bm x) \otimes \bm r_z(\bm z)\\
&=&(\bm r_x(\bm x)'\bm R_x^{-1}\otimes \bm r_z(\bm z)'\bm R_z^{-1})\bm r_x(\bm x) \otimes \bm r_z(\bm z)\\
&=&(\bm r_x(\bm x)'\bm R_x^{-1}\bm r_x(\bm x))\otimes (\bm r_z(\bm z)'\bm R_z^{-1}\bm r_z(\bm z))\\
&=&\bm r_x(\bm x)'\bm R_x^{-1}\bm r_x(\bm x)\bm r_z(\bm z)'\bm R_z^{-1}\bm r_z(\bm z).
\end{eqnarray*}
Thus,
\begin{eqnarray*}
  IMSE(\bm D) &=& \int_\mathcal{X} \int_\mathcal{Z} \{1-\bm r_x(\bm x)'\bm R_x^{-1}\bm r_x(\bm x)\bm r_z(\bm z)'\bm R_z^{-1}\bm r_z(\bm z)\} f^2(\bm z)/C_2 d\bm z d\bm x \\
   &=& 1-\int_\mathcal{X}\bm r_x(\bm x)'\bm R_x^{-1}\bm r_x(\bm x)d\bm x \int_\mathcal{Z}\bm r_z(\bm z)'\bm R_z^{-1}\bm r_z(\bm z)f^2(\bm z)/C_2 d\bm z\\
   &=& 1-(1-IMSE(\bm D_x))(1-IMSE(\bm D_z)).
\end{eqnarray*}
Thus, $\min_{\bm D} IMSE(\bm D)=1-(1-\min_{\bm Dx} IMSE(\bm D_x))(1-\min_{\bm D_z}IMSE(\bm D_z))$.
\qed
\bibliographystyle{ECA_jasa}
\bibliography{Ref_all}

\end{document}